\documentclass[12pt]{article}
\usepackage{epsfig,amssymb}

\newcommand{\bq}{\begin{equation}}
\newcommand{\eq}{\end{equation}}
\newcommand{\bqa}{\begin{eqnarray}}
\newcommand{\eqa}{\end{eqnarray}}
\newcommand{\ben}{\begin{enumerate}}
\newcommand{\een}{\end{enumerate}}
\newcommand{\bc}{\begin{center}}
\newcommand{\ec}{\end{center}}
\newcommand{\bqb}{\begin{eqnarray*}}
\newcommand{\eqb}{\end{eqnarray*}}

\def\llgm{\left\lgroup\matrix}
\def\rrgm{\right\rgroup}
\def\vectrl #1{\buildrel\leftrightarrow \over #1}
\def\partrl{\vectrl{\partial}}

\def\ie{{\it i.e.\/}}
\def\eg{{\it e.g.\/}}

\def\etal{{\it et.al.\/}}

\def\L{ {\cal L }}

\def\sw{s_W}
\def\cw{c_W}

\def\cwd{c_W^2}
\def\mw{m_W}

\def\mwd{m_W^2}

\def\mad{m_A^2}
\def\t{\hat t}
\def\s{\hat s}
\def\u{\hat u}

\def\Sn#1{\mathrm{Sign} #1 }

\def\xo{\tilde\chi_1}
\def\xt{\tilde \chi_2}
\def\tchi{\tilde \chi}
\def\cxi{\tilde \chi_i}

\def\ti{\tilde t_i}
\def\ton{\tilde t_1}
\def\tt{\tilde t_2}
\def\Stop{\tilde t}
\def\cfR{\cos\phi_R}
\def\cfL{\cos\phi_L}
\def\sfR{\sin\phi_R}
\def\sfL{\sin\phi_L}
\def\calpha{\cos \alpha}
\def\cbeta{\cos\beta}
\def\salpha{\sin \alpha}
\def\sbeta{\sin \beta}
\def\B{\tilde {\cal B}}
\def\Del{\tilde { \Delta}}

\def\pr#1#2#3{ Phys. Rev. ${\bf{#1}}$, #2 (#3)}

\def\pl#1#2#3{ Phys. Lett. ${\bf{#1}}$, #2 (#3)}
\def\prep#1#2#3{ Phys. Rep. ${\bf{#1}}$, #2 (#3)}

\def\np#1#2#3{ Nucl. Phys. ${\bf{#1}}$, #2 (#3)}
\def\zp#1#2#3{ Z. f. Phys. ${\bf{#1}}$, #2 (#3)}
\def\epj#1#2#3{ Eur. Phys. J. ${\bf{#1}}$, #2 (#3)}

\def\fortp#1#2#3{ Fortsch. Phys. ${\bf{#1}}$, #2 (#3)}
\def\cpc#1#2#3{Comput. Phys. Commun. ${\bf{#1}}$, #2 (#3)}

\def\JPhysG#1#2#3{J. Phys. G ${\bf{#1}}$, #2 (#3)}
\def\AnnPhys#1#2#3{Ann. Phys. (N.Y.)${\bf{#1}}$, #2 (#3)}

\begin{document}
\pagenumbering{arabic}
\thispagestyle{empty}
\def\thefootnote{\fnsymbol{footnote}}
\setcounter{footnote}{1}

\begin{flushright}
 THES-TP 2000/07
\\ hep-ph/0007110.
\\ May 2000
\end{flushright}
\vspace{2cm}
\begin{center}
{\Large\bf The $\gamma \gamma \to A^0 A^0 $ process at a $\gamma
\gamma $ Collider.}\footnote{Partially supported by the European
Community grant ERBFMRX-CT96-0090.}
 \vspace{1.5cm}  \\
{\large G.J. Gounaris, P.I. Porfyriadis }\\ \vspace{0.7cm}
Department of Theoretical Physics, Aristotle University of
Thessaloniki,\\ Gr-54006, Thessaloniki, Greece.\\

\vspace{0.2cm}

\vspace*{1cm}

{\bf Abstract}
\end{center}
The  helicity amplitudes for the process
$\gamma \gamma \to A^0 A^0$ are studied to 1-loop
order in the minimal SUSY (MSSM) model,
 where $A^0$ is the CP-odd Higgs particle.
  Simple exact analytic formulae are
obtained, in terms of the $C_0$ and $D_0$ Passarino-Veltman
functions; in spite of the fact that
the  loop-diagrams often involve  different particles
running along their sides.
For a usual  mSUGRA set of parameters,
$\sigma (\gamma \gamma \to A^0 A^0) \sim (0.1-0.2)\rm fb $
is expected. If SUSY is realized in Nature,
 these  expressions should   be useful for
understanding the Higgs sector. \par

\def\thefootnote{\arabic{footnote}}
\setcounter{footnote}{0}
\clearpage

\section{Introduction}

\hspace{0.7cm} If  in the future
$e^-e^+$ Linear Colliders (LC) \cite{LC},
 the option to develop high energy $\gamma \gamma$
collisions  will also be available, then many
new opportunities for new physics (NP) searches
should arise. Employing back-scattering of laser photons,
this option transforms
an\footnote{In this case it would be best to run LC
in its   $e^-e^-$ mode, \cite{gamma2000, Telnov-gg2000}.} LC
to essentially a $\gamma \gamma $ Collider
($\rm LC_{\gamma \gamma}$) with  about $80\%$ of the
initial energy and a comparable luminosity
\cite{gamma2000, Telnov-gg2000}.
 The importance of $\rm LC_{\gamma \gamma}$ stems from the
fact that  the cross sections
for gauge boson and top production in $\gamma \gamma $ collisions
at sufficiently high energies, are often considerably  larger
than the corresponding quantities in  the
$e^-e^+$ case \cite{LCgg, LCgg-th1}.  \par

To some extent, such an enhancement should  arise
for Higgs production also.
  For the neutral Higgs particles in particular, an
$\rm LC_{\gamma \gamma}$ may act as a Higgs factory which can
be used to study their  detail properties, including  possible
Higgs  anomalous couplings
\cite{Higgs-factory, ggVV-anomalous}. Since the anomalous
gauge boson, top and Higgs couplings  are interconnected
and constitute an important possible  source of new physics,
an $\rm LC_{\gamma \gamma}$ should be very helpful for its
identification. In case the NP scale is very high, such forms of
NP may be described by the complete list of $dim=6$ operators
involving gauge bosons and/or quarks of the third
family presented in \cite{ top-anomalous}. \par

Alternatively, it may turn out that the NP scale is
nearby,  as it would be expected in the usual  SUSY scenario
\cite{SUSY-reviews}. In such a case many neutral
spinless  particles of Higgs and sneutrino type may exist,
and an  $\rm LC_{\gamma \gamma}$
may be used for an  s-channel production of the
CP-even light and   heavy neutral Higgs bosons
$h^0$ and $H^0$ respectively, as well as the CP-odd $A^0$.
The study
of the various branching ratios, and   the
polarization of the incoming photons, could then be
very helpful to establish and
disentangle the nature of these Higgs particles
\cite{single-H0A0, Spira-Higgs}.\par

Once any  of these spinless bosons is discovered, its
properties should be carefully  looked at,
in order to be sure that
they  fulfill the SUSY expectations. Motivated by this, we study in
this paper the process $\gamma \gamma \to A^0 A^0$ in the context
of a  minimal SUSY model, where no new sources of CP violation,
apart from those already known in the
 Standard Model (SM) Yukawa potential, are assumed to
exist. Thus, the various new SUSY couplings are taken to be real,
but no specific assumption on their relative magnitudes
or signs is made \cite{Djouadi-SUSY-group}.
As we will see below, in such a
case, there are only two independent helicity amplitudes for
$\gamma \gamma \to A^0 A^0$ , denoted below  as $F_{++}$ and
$F_{+-}$, where the  indices describe the helicities of the
incoming photons. \par

It is also interesting to study the phases of these amplitudes.
The motivation for this stems from the recent observation
in \cite{gg-gZ, ggZZ}, that at c.m energies
$\gtrsim 250GeV$, out of the many
independent helicity  amplitudes  for the processes
$(\gamma \gamma \to \gamma \gamma,~ \gamma Z,~ ZZ)$,
only the two helicity conserving amplitudes
$F_{++++}$ and $F_{+-+-}$ are important, which moreover turn out
to be almost purely imaginary\footnote{For $\gamma \gamma \to ZZ$ in SM the further
assumption is made that the standard Higgs particle
is light; \eg~ below $\sim 200 GeV$.}.
The physical reason for this result is not very
clear \cite{gg-gZ, ggZZ}. Therefore,
it seems worthwhile to investigate what happens
in other  processes, like \eg~ the
neutral Higgs boson production, which, as
the neutral gauge bosn production,
  also  vanish at tree order and they
first appear at the 1-loop level.\par

Below, in Section 2 we give an  overall view of the $\gamma \gamma
\to A^0A^0$ helicity amplitudes in SUSY. The  needed
SUSY vertices  appear in Appendix A, while the corresponding
contributions to the amplitudes are given in Appendix B.
The results are expressed in terms of $C_0$ and $D_0$
Passarino-Veltman functions only \cite{Passarino},
 using expressions analogous to those encountered in the
$\gamma \gamma \to ZZ$ calculation \cite{ggZZ}.
Finally in Section 3, we give our Conclusions.\par

Coming now to the related studies already existing in the
literature, we first remark that $\gamma \gamma \to h^0 h^0 $ has
been studied in SM by Jikia \cite{Jikia-Higgs}. In the non-linear
gauge defined in (\ref{gauge-fixing-L}) and used here,
the only contributing
diagrams involve $W$ or top-loops, similar to those appearing
in Figs.\ref{WH-diag},\ref{top-diag}. We have repeated the
calculations of \cite{Jikia-Higgs}
and agree with the results,
apart from the overall sign of the\footnote{For the gauge boson
polarization vectors, here and in \cite{gg-gZ, ggZZ},
we use the same conventions as in \cite{Jikia-ggZZ}.
The only difference is that we use the JW convention
\cite{Jacob}, which  introduces an additional minus
to the polarization vector of a longitudinal
"Number 2" Z and affects
$\gamma \gamma \to ZZ,~ \gamma Z$.}  $F_{++}$ amplitude.
For the top contributions, our results are fully consistent with
those of \cite{Glover-Higgs}.
The relevant amplitudes are presented
and compared to  those of $\gamma \gamma \to A^0A^0$
at the end of Section 2. \par

In \cite{Zhu-SUSY-h0} a calculation  of $\gamma \gamma \to h^0
h^0$  in a  general SUSY model has been presented in terms of the
general $C_j$ and $D_j$ of Passarino and Veltman
functions. The production of two neutral Higgs pairs in SUSY models at
hadronic Colliders has also been studied in
\cite{Drees-LHC-SUSY-Higgs}; where of course the complications
from loops involving $W$-bosons, or "single" and  "mixed"
charginos, are avoided. Finally the  processes
$\gamma \gamma \to H^0 H^0$ and $\gamma \gamma \to A^0 A^0$
have also appeared in a non-Supersymmetric
gauge model involving a two Higgs doublet scalar sector
  \cite{twoHiggs-ggH0H0, twoHiggs-ggA0A0}.

\section{An overall view of the $\gamma \gamma \to A^0 A^0 $
amplitudes.}

The invariant helicity amplitudes for\footnote{We use the same
conventions as in \cite{gg-gZ, ggZZ}.} $\gamma \gamma \to A^0A^0$
are denoted as $F_{\lambda_1, \lambda_2}(\s, \t, \u)$, where
$\lambda_j$ describe the helicities of the incoming photons, and
the kinematics are defined in Appendix B. Assuming that the SUSY
Higgs potential is CP-invariant we get (see (\ref{CP-constraint}))
\bq
F_{\lambda_1, \lambda_2 }(\s,\t,\u)=
F_{-\lambda_1, -\lambda_2 }(\s,\t,\u) ~~~,
\label{CP-constraint-text}
\eq
which implies that there are only two independent helicity
amplitudes, $F_{++}(\s, \t, \u)$ and $F_{+-}(\s, \t, \u)$.

As in \cite{gg-gZ, ggZZ}, we employ the non-linear gauge
of \cite{Dicus}, which implies   the gauge fixing and FP-ghost
interactions of (\ref{gauge-fixing-L}, \ref{ghost-L}), leading to
the conclusion that there are  no
$\gamma W^\pm G^\mp$,  $Z W^\pm G^\mp$ vertices.
The diagrams contributing to $A^0$-pair production are then
given in Figs.\ref{WH-diag}-\ref{stop-diag}.

The contribution to the $F_{++}$ and $F_{+-}$
amplitudes from the diagrams in Fig.\ref{WH-diag}
consists of two types. The first is induced by the
two diagrams in the first line  in Fig.\ref{WH-diag} and describes
the $(h^0, ~H^0)$-pole contributions appearing in
(\ref{FppWH-h-pole}, \ref{FppWH-H-pole}).
The diagrams in the second to last line of Fig.\ref{WH-diag}
involve loops in which
$W^\pm$ and/or  $H^\pm$ are running along their  internal lines.
These induce the second type of contributions
contained in (\ref{FppWH}, \ref{FpmWH}),
  and  expressed in terms  of the $(C_0,~ D_0)$-functions
explained in
(\ref{C0s}-\ref{DAAut}); as well as  the functions
$\tilde F^{WH^\pm}$, $\tilde F^{H^\pm W}$, $E_1^{WH^\pm}$
defined in (\ref{Fstu}, \ref{E1st}).
The contributions
(\ref{FppWH}, \ref{FpmWH}) give the largest effect
to the $\gamma \gamma \to A^0A^0$ amplitudes,  for
the numerical applications considered below. \par

The chargino loop contribution is described by the diagrams in
Fig.\ref{chargino-diag}. It  consists also  of an $(h^0,~ H^0)$-pole
contribution given in (\ref{Fppchi-pole}); the box contributions
involving a "single chargino"-loop giving
(\ref{Fppchi}, \ref{Fpmchi}); and the "mixed chargino"
contribution (\ref{Fppchi1chi2}, \ref{Fpmchi1chi2}),
arising   when both charginos are running along the loop.
Analytically, the later is the most complicated one.
Nevertheless, it is simple enough to be possible to write it.
Numerically, it has to be taken into
account only when both charginos are relatively
light. \par

The $t$ and $b$ quark contributions are described by the diagrams
in Fig.\ref{top-diag}. They are given in (\ref{Fpptop-pole}) for
the $(h^0,~ H^0)$-pole contribution, and
in (\ref{Fpptop}, \ref{Fpmtop}) for the box diagrams.\par

As an example of a sfermion contribution, we only considered the
one arising from the $(\Stop_1,~ \Stop_2)$-loop,
described by the diagrams in Fig.\ref{stop-diag}. Their
contributions are  given by
(\ref{Fppstop-pole}-\ref{Fpmstop}).\par

\begin{table}[hbt]
\begin{center}
{ Table 1: mSUGRA parameters in
Figs. \ref{ggAA-fig-mSUGRA1}-\ref{ggAA-fig-mSUGRA3}.
\cite{mSUGRA-param}.}\\
\vspace*{0.3cm}
\begin{tabular}{||c|c|c|c||}
\hline \hline
\multicolumn{1}{||c|}{}&
\multicolumn{1}{|c|}{mSUGRA(1)}&
\multicolumn{1}{|c|}{mSUGRA(2)}&
\multicolumn{1}{|c||}{mSUGRA(3) (light stop)}
\\ \hline
 $\tan\beta $ & 3  & 30  & 3 \\ \hline
\hline
\multicolumn{4}{||c||}{at the Unification scale}
\\ \hline
 $m_0(\rm GeV)$  & 100  & 160 & 100 \\ \hline
$M_{1/2}(\rm GeV)$  & 200  & 200 & 200 \\ \hline
$A_0(\rm GeV) $  & 0  & 600 & -715 \\ \hline
$\rm sign(\mu)$   & + & + & + \\ \hline \hline
\multicolumn{4}{||c||}{at the Electroweak  scale}
\\ \hline
 $M_2(\rm GeV)$  & 152  & 150 & 153 \\ \hline
$\mu (\rm GeV)$  & 316  & 263  & 435 \\ \hline
$m_{A^0}(\rm GeV)$  & 375  & 257 & 489 \\ \hline
$m_{h^0}(\rm GeV)$  & 97.7  & 108 & 101 \\ \hline
$m_{H^0}(\rm GeV)$  & 379  & 257 & 490 \\ \hline
$m_{H^\pm}(\rm GeV)$  & 383  & 269 & 495 \\ \hline
$A_t (\rm GeV) $  & -373  & -258 & -500 \\ \hline
$m_{\tilde \chi_1^+} (\rm GeV)$  & 128  & 132  & 138 \\ \hline
$m_{\tilde \chi_2^+}(\rm GeV)$  & 346  & 295 & 454 \\ \hline
$m_{\tilde t_1} (\rm GeV)$  & 295.4 & 353   & 133 \\ \hline
$m_{\tilde t_2}(\rm GeV)$  & 494.2   & 469  & 491 \\ \hline \hline
\end{tabular}
\end{center}
\end{table}

\begin{figure}[p]
\vspace*{-3cm}
\[
\epsfig{file=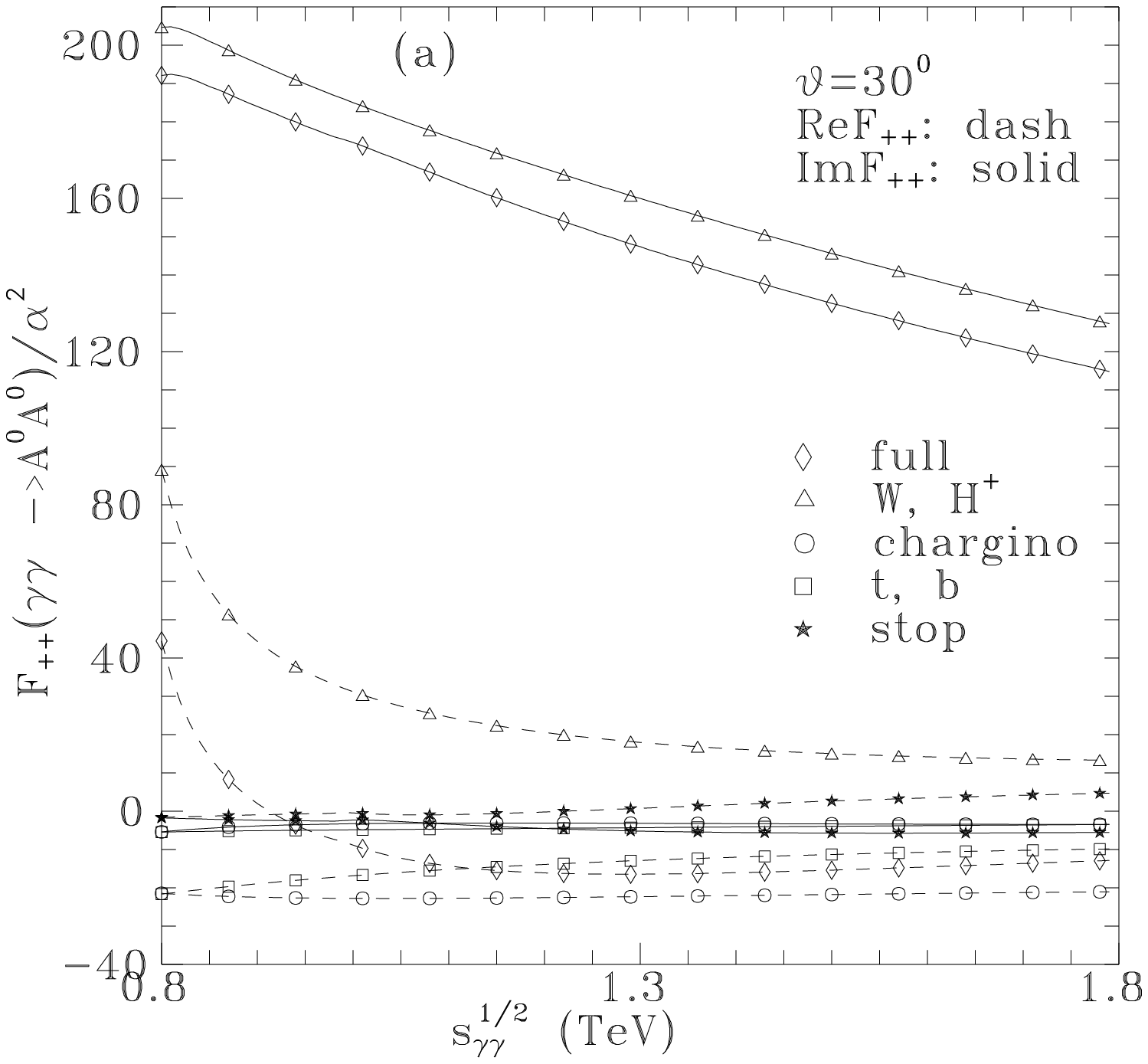,height=7.5cm}\hspace{0.5cm}
\epsfig{file=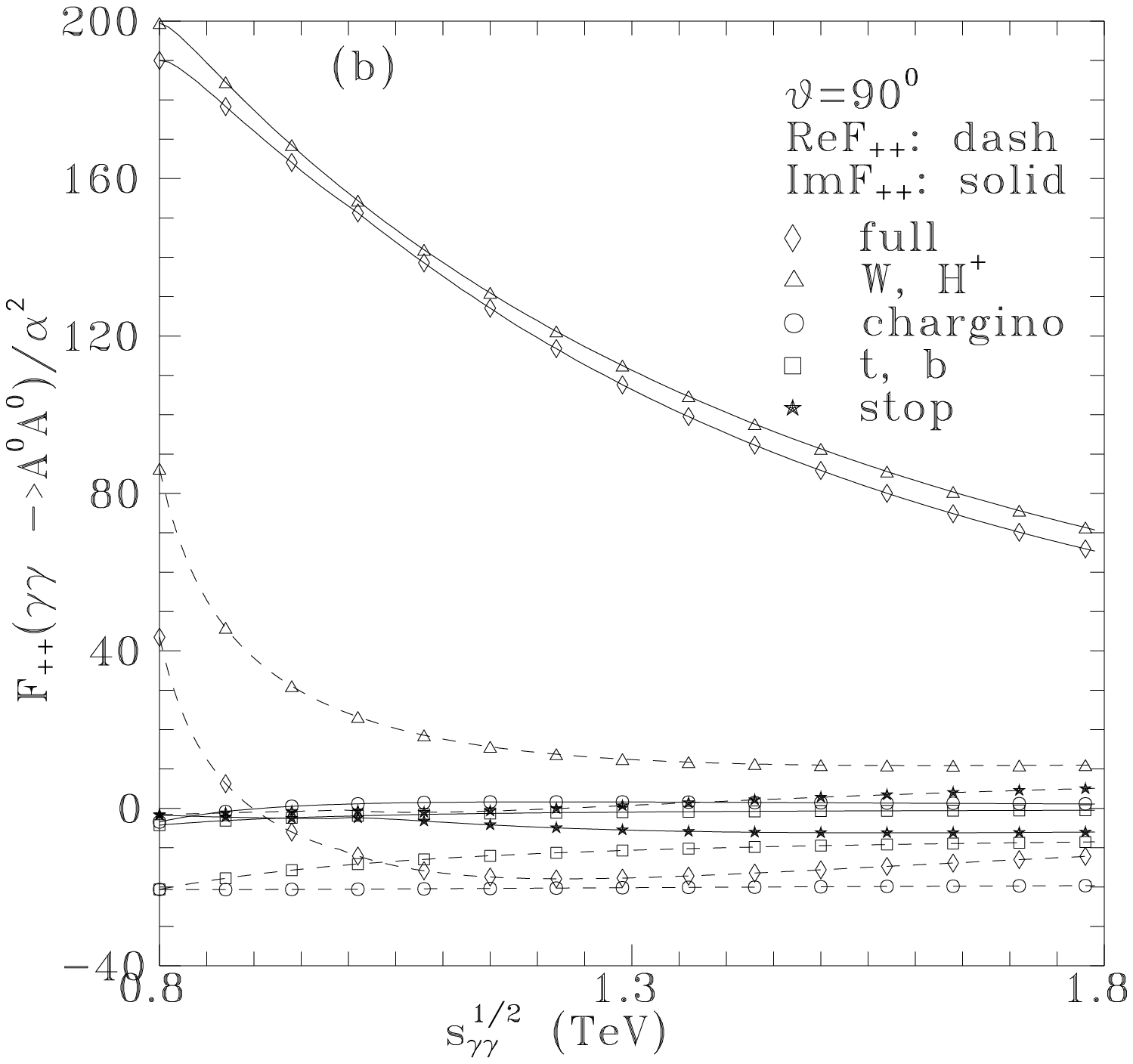,height=7.5cm}
\]
\vspace*{0.5cm}
\[
\epsfig{file=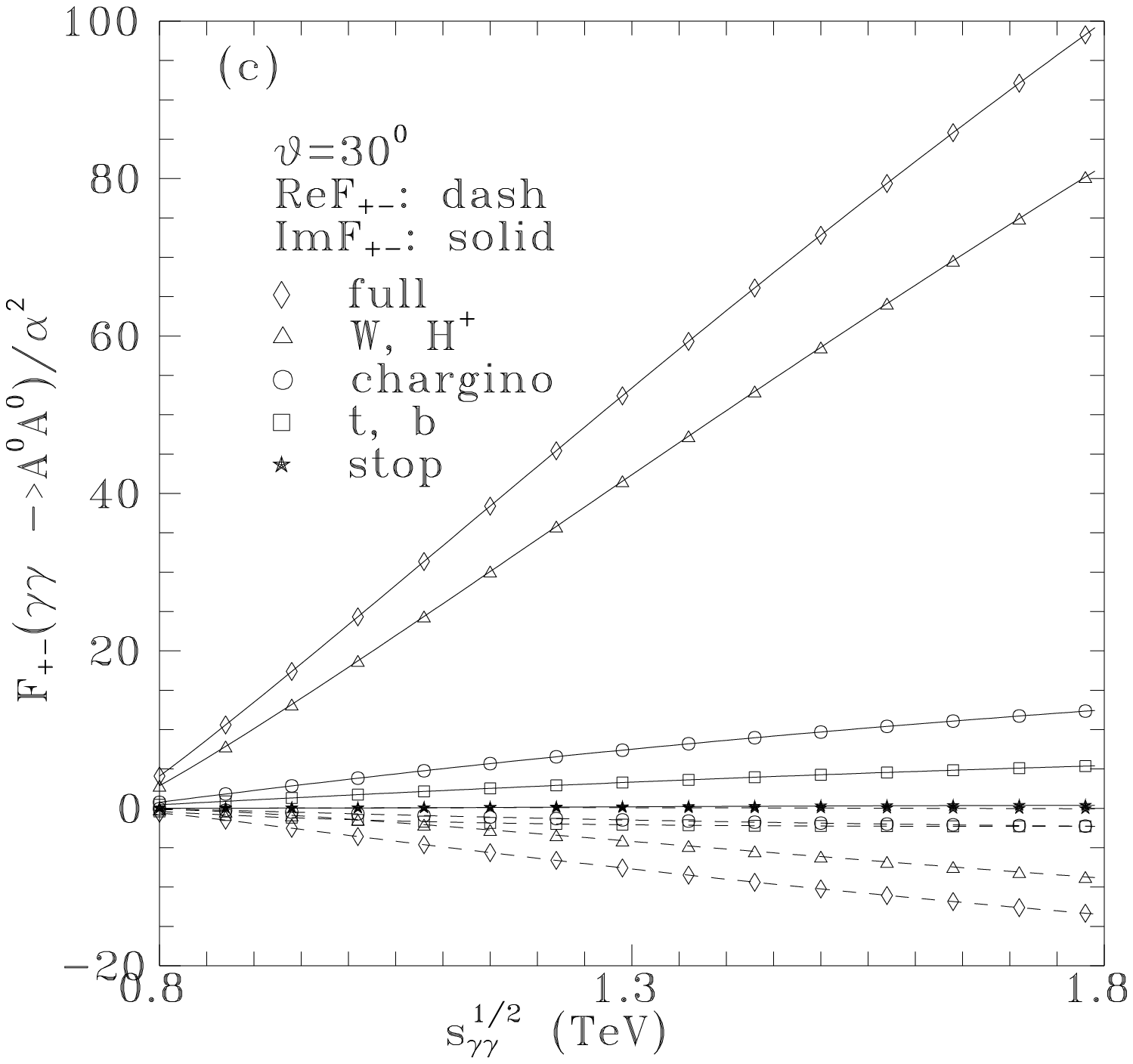,height=7.5cm}\hspace{0.5cm}
\epsfig{file=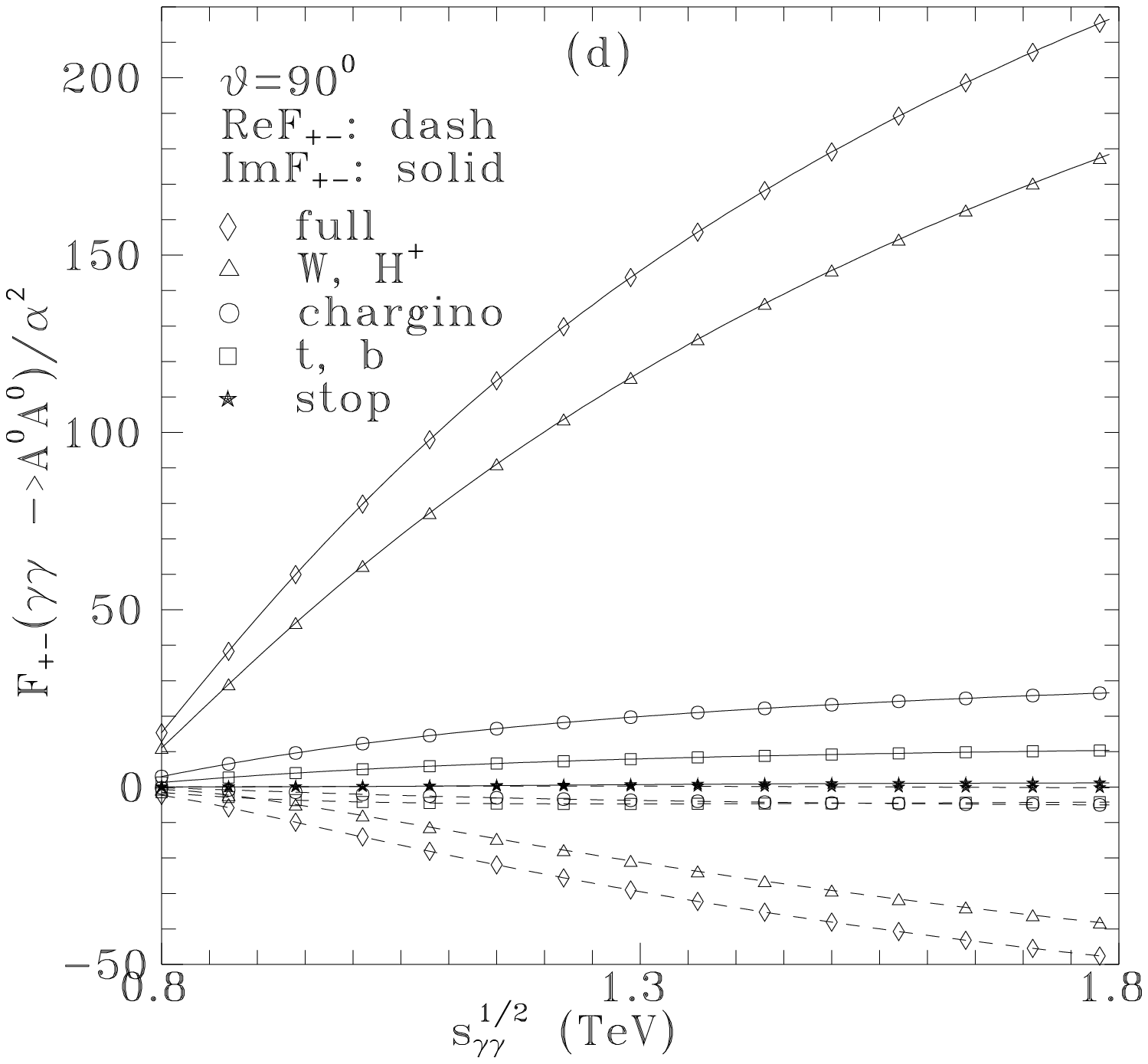,height=7.5cm}
\]
\caption[1]{$\gamma \gamma \to A^0A^0$
helicity amplitudes as functions of the
$\gamma \gamma $ center-of-mass energy
$s_{\gamma \gamma}\equiv \s$ for
mSUGRA(1); see Table 1.}
\label{ggAA-fig-mSUGRA1}
\end{figure}

\begin{figure}[p]
\vspace*{-3cm}
\[
\epsfig{file=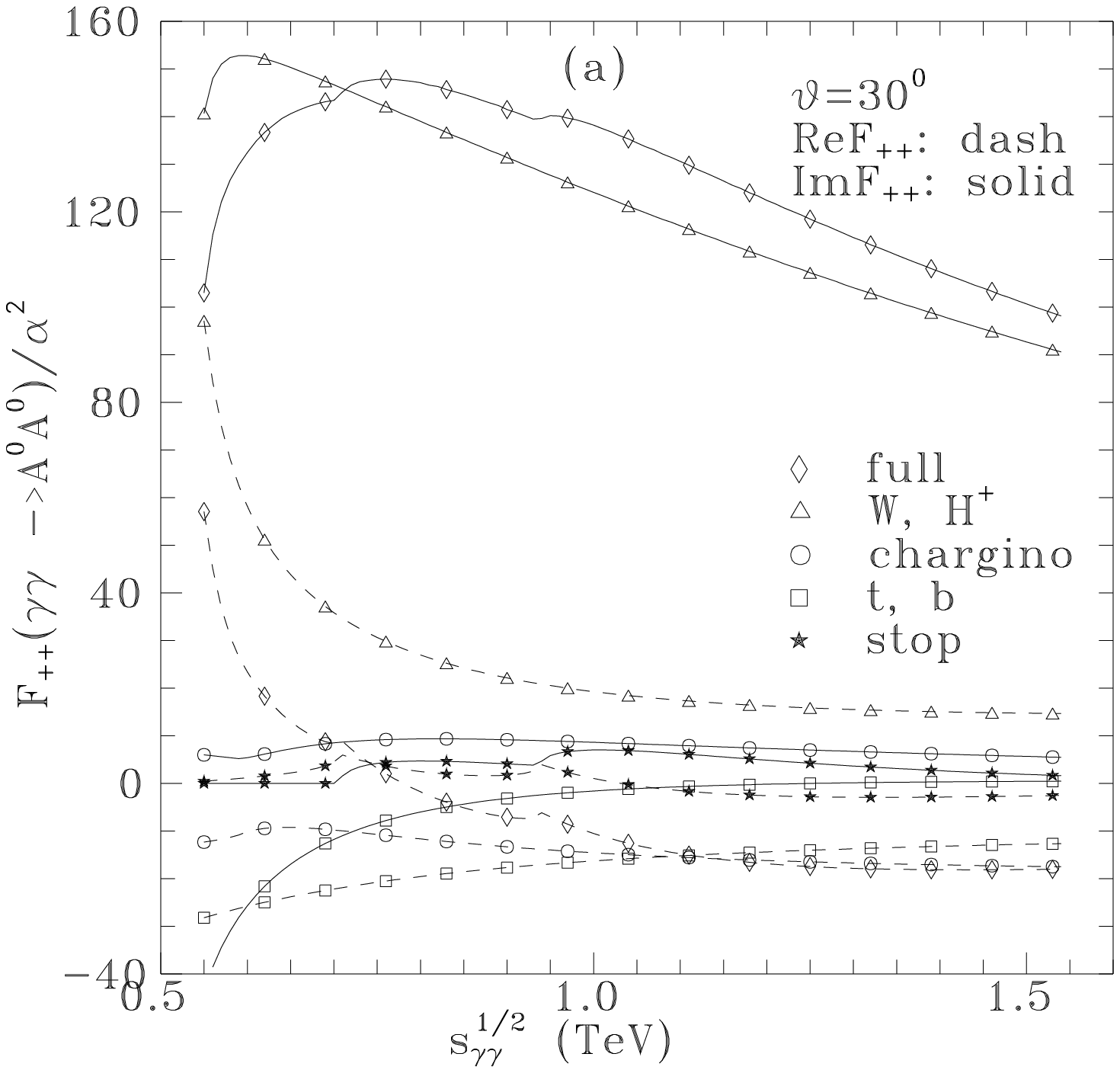,height=7.5cm}\hspace{0.5cm}
\epsfig{file=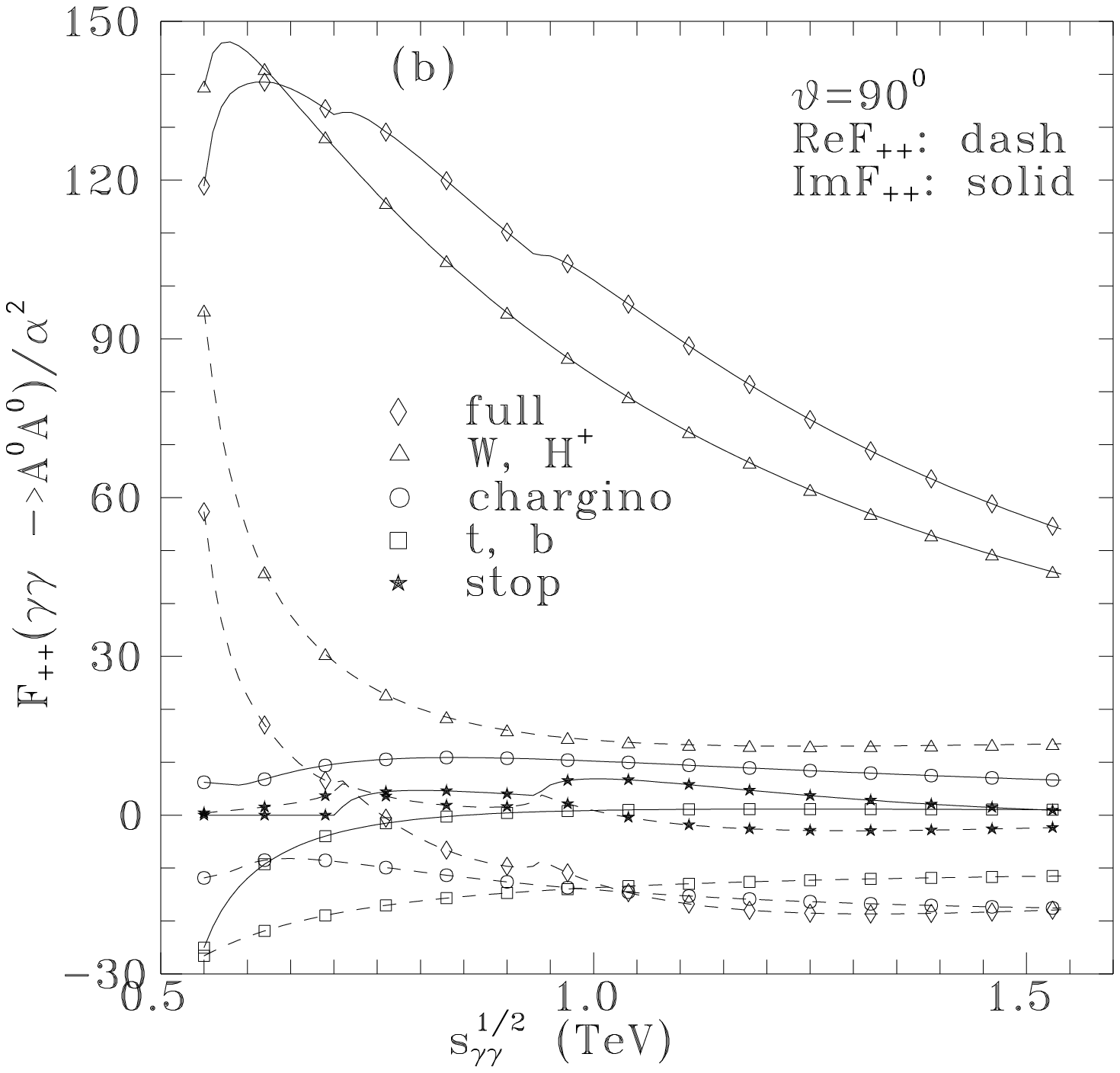,height=7.5cm}
\]
\vspace*{0.5cm}
\[
\epsfig{file=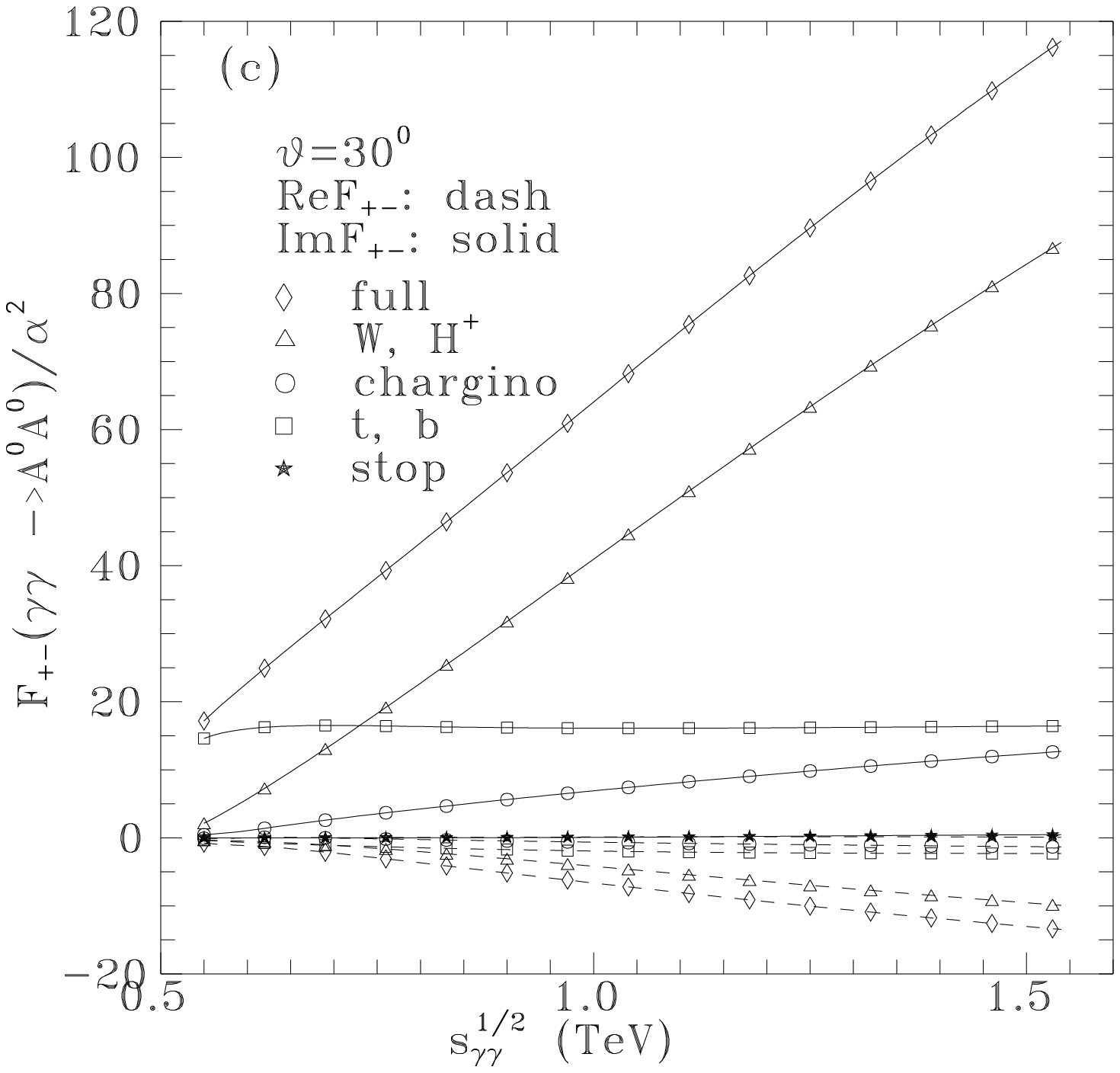,height=7.5cm}\hspace{0.5cm}
\epsfig{file=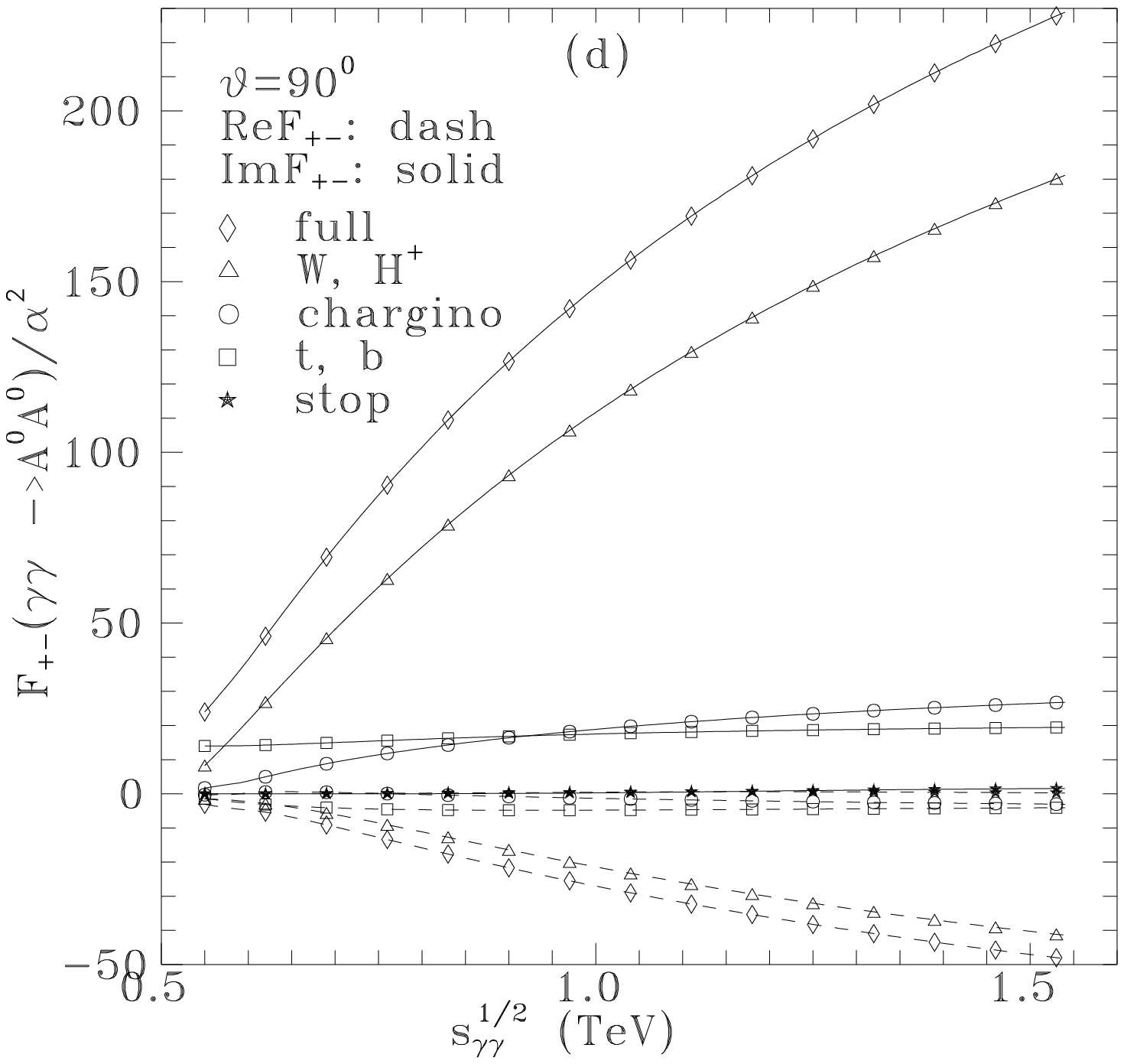,height=7.5cm}
\]
\caption[1]{$\gamma \gamma \to A^0A^0$
helicity amplitudes as functions of the
$\gamma \gamma $ center-of-mass energy
$s_{\gamma \gamma}\equiv \s$ for  mSUGRA(2);
see Table 1.}
\label{ggAA-fig-mSUGRA2}
\end{figure}

\begin{figure}[p]
\vspace*{-3cm}
\[
\epsfig{file=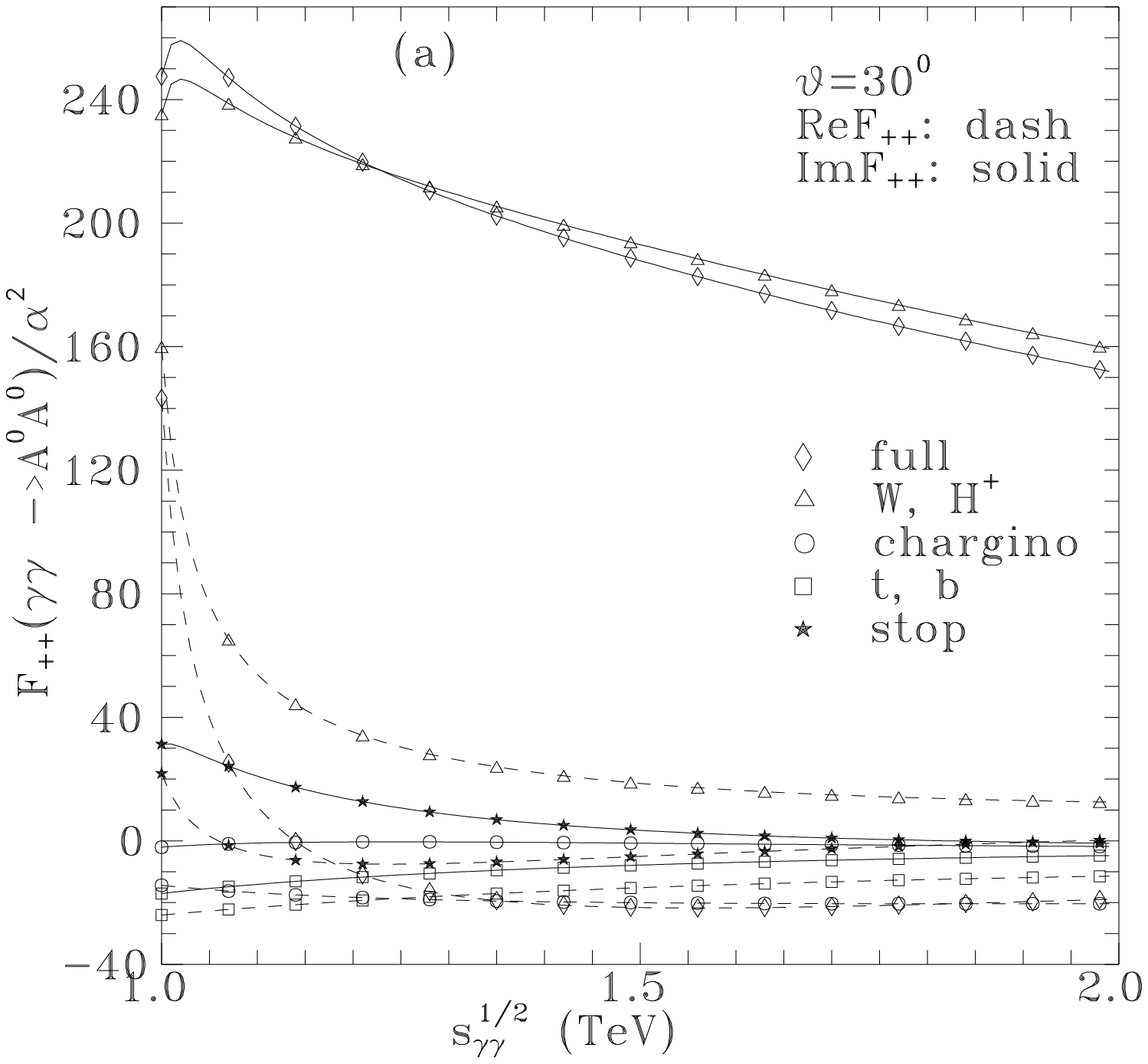,height=7.5cm}\hspace{0.5cm}
\epsfig{file=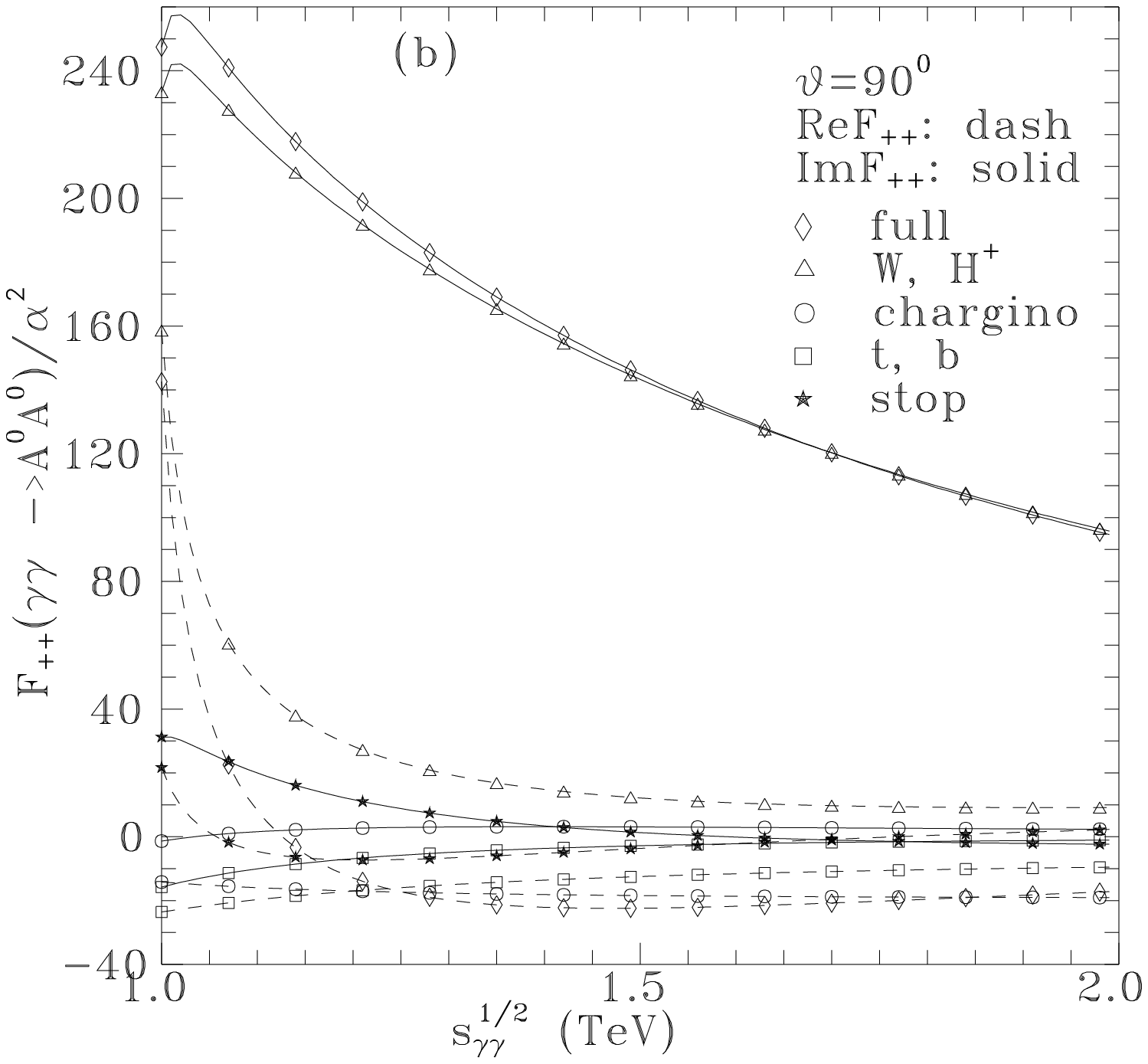,height=7.5cm}
\]
\vspace*{0.5cm}
\[
\epsfig{file=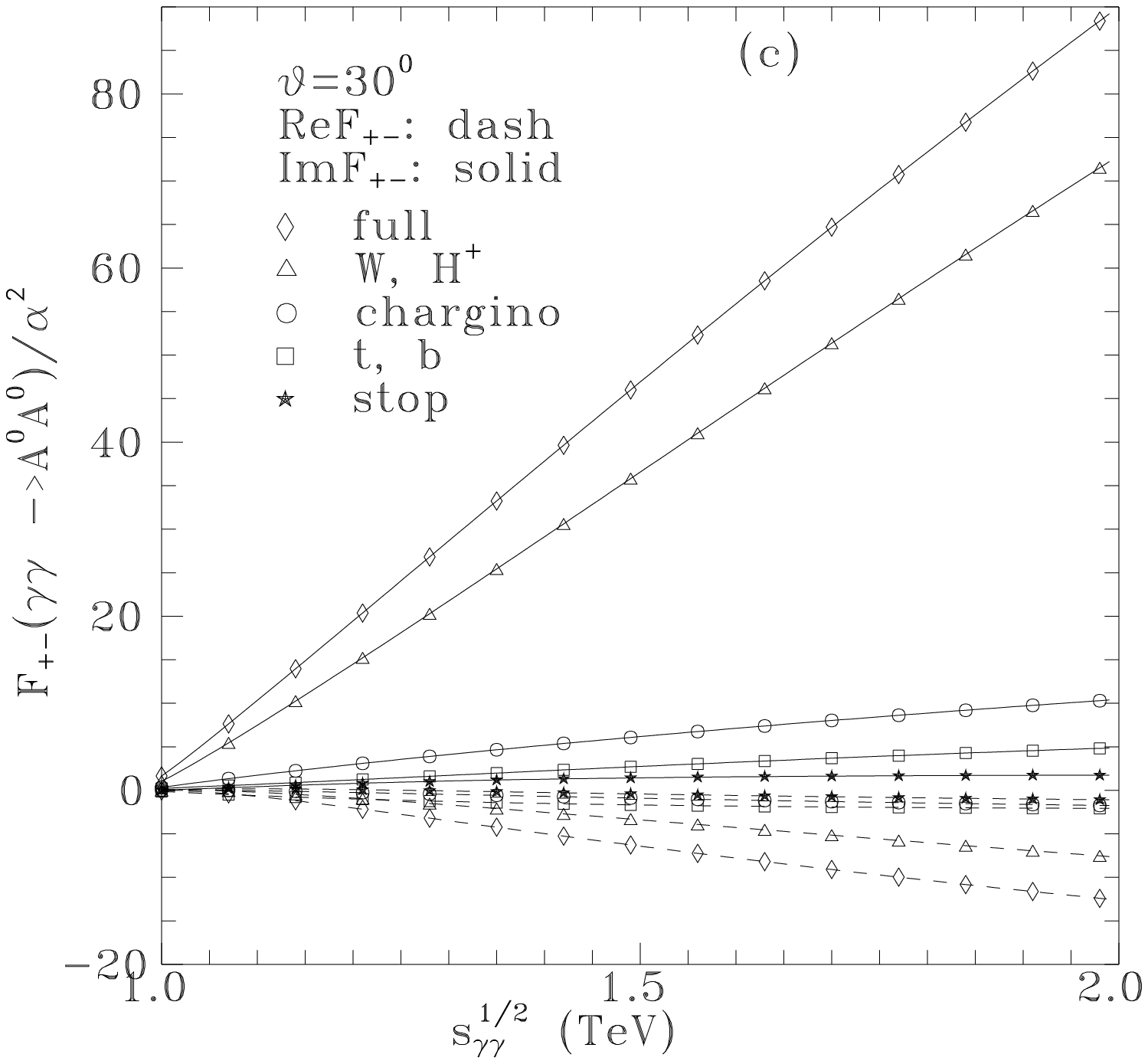,height=7.5cm}\hspace{0.5cm}
\epsfig{file=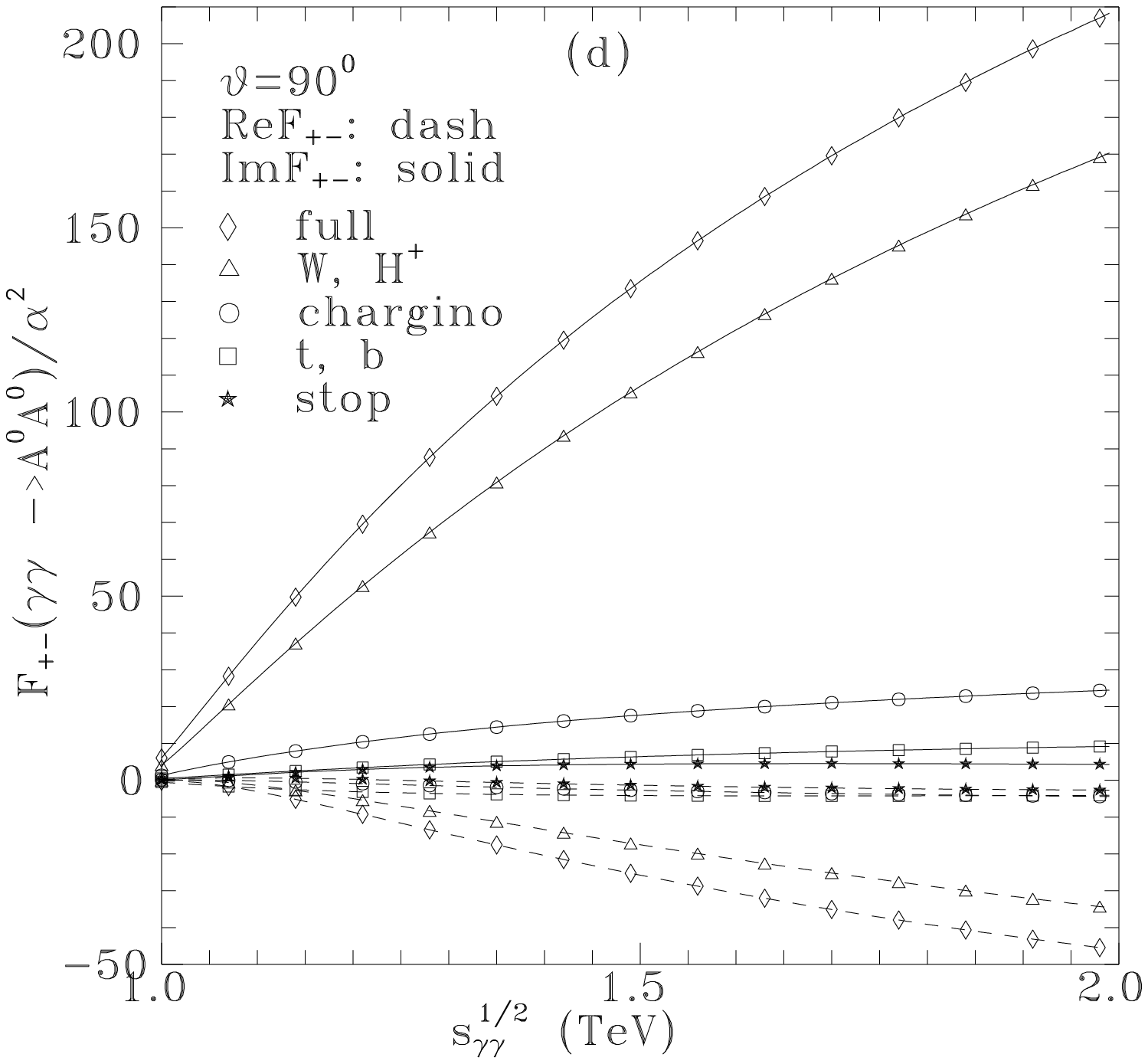,height=7.5cm}
\]
\caption[1]{$\gamma \gamma \to A^0A^0$
helicity amplitudes as functions of the
$\gamma \gamma $ center-of-mass energy
$s_{\gamma \gamma}\equiv \s$ for mSUGRA(3);
see Table 1.}
\label{ggAA-fig-mSUGRA3}
\end{figure}

For the numerical applications  we use the three CP-invariant
mSUGRA set of parameters introduced in
\cite{mSUGRA-param, Djouadi-SUSY-group}
and presented in Table 1. For the electromagnetic coupling
we take $\alpha=\rm 1/127.8 $. The  results are shown in
Figs.\ref{ggAA-fig-mSUGRA1}-\ref{ggAA-fig-mSUGRA3}.

The real and imaginary parts of the
helicity amplitudes $F_{++}(\gamma \gamma \to A^0 A^0)$ and
$F_{+-}$ are presented in
Figs. \ref{ggAA-fig-mSUGRA1}-\ref{ggAA-fig-mSUGRA3}
\cite{Oldenborgh}.
As indicated there, the most important contributions to the
amplitudes arise from the $(W,~H^\pm)$-loop diagrams presented in
the 2nd to last line of Fig.\ref{WH-diag} and appearing in
(\ref{FppWH}, \ref{FpmWH}). At sufficiently high energies,
these contributions are mainly imaginary.
But the predominance of the imaginary parts of
the amplitudes is not so strong, as the one observed in the
gauge boson production cases \cite{gg-gZ, ggZZ}.\par

As indicated in
Figs. \ref{ggAA-fig-mSUGRA1}-\ref{ggAA-fig-mSUGRA3}, the
chargino contribution is generally quite important;
while   the  $t,~b$-quark contribution is somewhat
smaller;  and  the stop contribution is negligible for
the above cases.\par

For the $t,~ b$-quark contribution we also remark that
in the mSUGRA(1) and  mSUGRA(3) cases, where
$\tan \beta$ is small, the $b$-contribution is negligible
compared to the top one.
On the contrary, for the mSUGRA(2) case of $\tan\beta=30$,
the $b$-quark contribution may be more important than the
$t$-quark one.\par

\begin{figure}[p]
\vspace*{-2cm}
\[
\epsfig{file=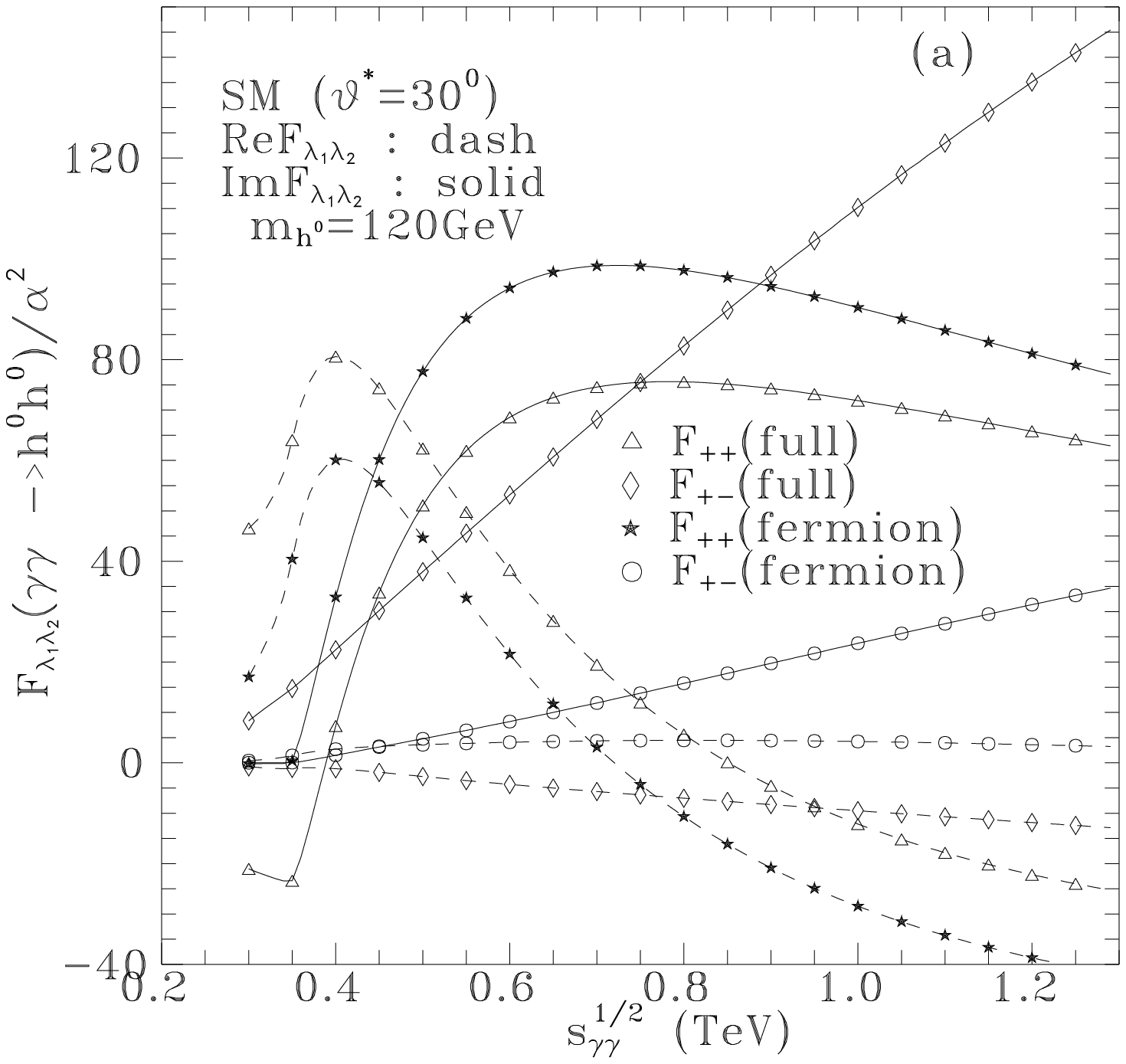,height=7.5cm}\hspace{0.5cm}
\epsfig{file=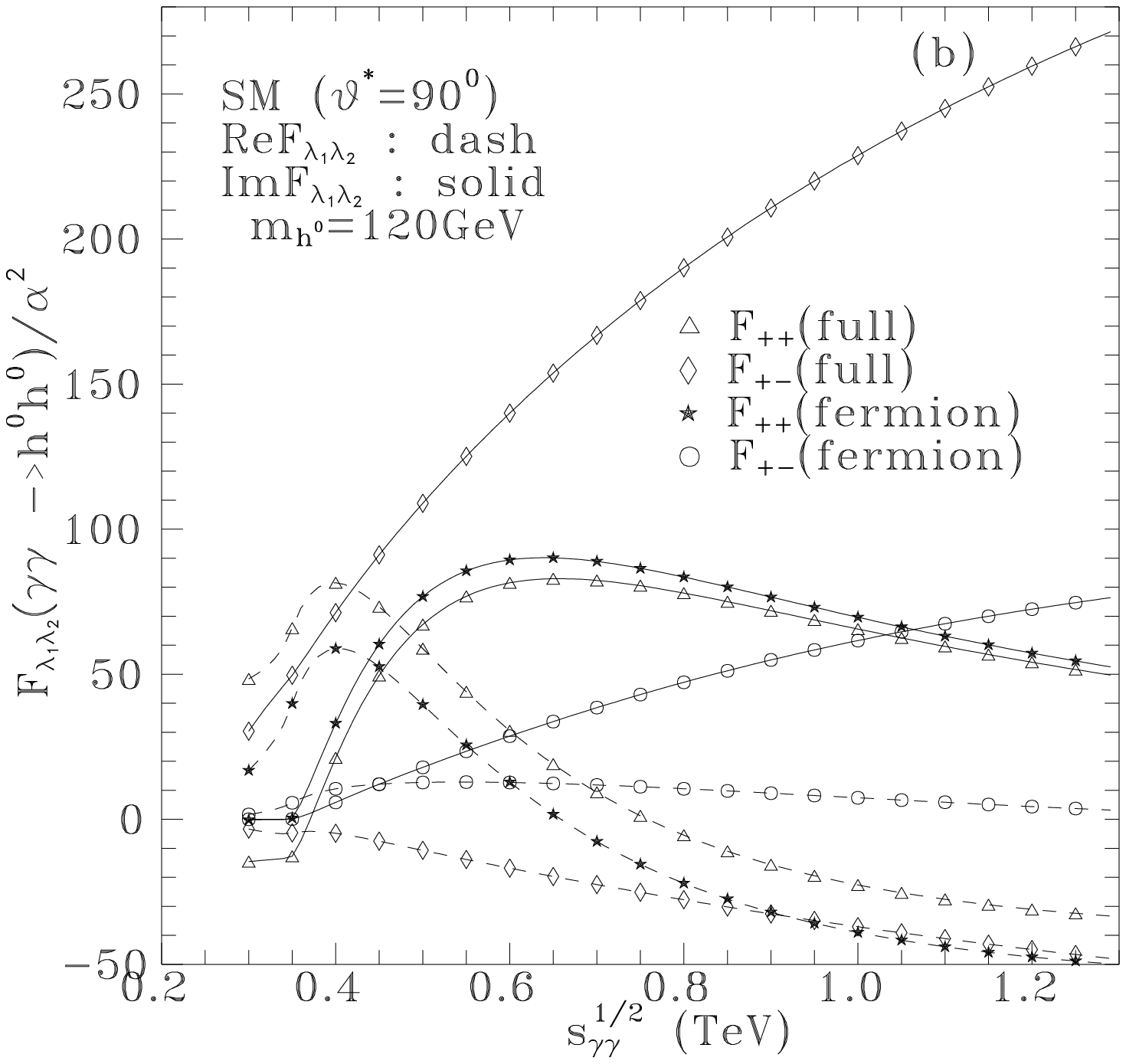,height=7.5cm}
\]
\caption[1]{$\gamma \gamma \to h^0 h^0$
helicity amplitudes as functions of the
$\gamma \gamma $ center-of-mass energy
$s_{\gamma \gamma}\equiv \s$ in SM. The SM fermion
contribution is also separately given.}
\label{ggHH-fig-SM}
\end{figure}

For comparison, we have also looked at the $F_{++}$ and $F_{+-}$
amplitudes for $\gamma \gamma \to h^0 h^0$ in the Standard Model.
The results for $m_{h^0}=120 \rm GeV$ are given in
Fig.\ref{ggHH-fig-SM}. For the $F_{++}$ we find that the
$top$-loop contributions is comparable the $W$-one, and the
amplitude is never particularly imaginary. It is only for
$F_{+-}$, for which there is no Higgs-pole contribution;
that at energies  $\gtrsim 600 \rm GeV$,
the $W$-loop is more important than the top-one,
and the imaginary part of the  amplitude becomes predominant.

\begin{figure}[hbt]
\vspace*{0cm}
\[
\epsfig{file=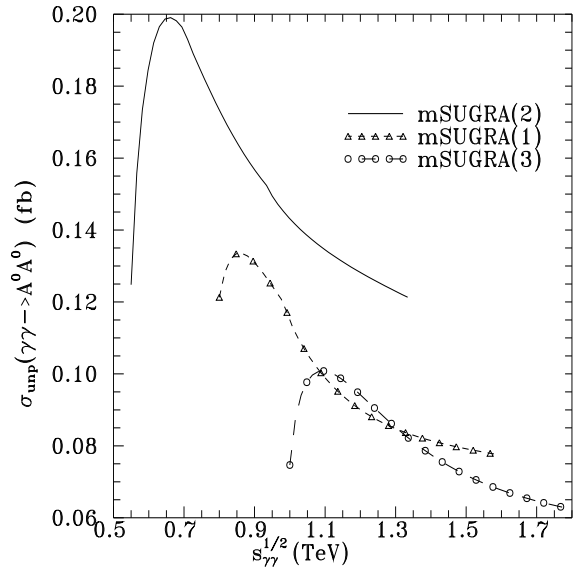,height=7.5cm,width=7.5cm}
\]
\caption[1]{Unpolarized total cross section for
$\gamma \gamma \to A^0 A^0$ in the region
$30^0 < \vartheta^* <150^0$ for the  SUSY models
SUGRA(1), SUGRA(2) and SUGRA(3); see Table 1.}
\label{sig-ggAA-fig}
\end{figure}

The $\gamma \gamma \to A^0 A^0$ unpolarized
cross section  for the
sets of parameters in Table 1, are given in Fig.\ref{sig-ggAA-fig}.
It lies in the range of $\sim (0.1-0.2)\rm fb$, which is similar
but somewhat smaller, than the result expected for
$\sigma (\gamma \gamma \to h^0 h^0)$ in
SM for $m_{h^0}\sim 120 \rm GeV$ \cite{Jikia-Higgs}.
This result does not seem
particularly sensitive to SUSY parameters like \eg~ $\tan
\beta$; but mainly depends on the $A^0$-mass. It should also
be compared to the situation for a single $A^0$ or $H^0$ production
studied in \cite{Barger}.  We also remark that a
  cross section at the $(0.1-0.2)\rm fb$-level may
be observable, if a luminosity of \eg~ $\L_{\gamma
\gamma} \sim 250 \rm fb^{-1}/year $ is realized  in TESLA
\cite{gamma2000, Telnov-gg2000}.\par

\section{Conclusions}

The Higgs sector, which is responsible for giving masses to
almost all particles immediately after our Universe started, is
definitely the most fascinating part of the present
elementary particle theory. Motivated by this
and assuming that the SUSY
option is chosen by nature, we have studied here the process
$\gamma \gamma \to A^0 A^0$.

In the non-linear gauge used here, the types of contributing
diagrams may be divided into two categories
constructed on the basis of whether
an s-channel neutral  Higgs-pole is involved or not.
Each category may then be
further divided into three classes, on the basis of whether
their loops involve
the $(W,~ H^\pm)$-pair, charginos or sfermions.
General formulae have been presented which allow the
description of the process in any SUSY model, minimal or
non-minimal.

For the numerical applications we only considered three SUGRA
examples presented in Table 1, leading to an $A^0$ heavier than
$\sim 250\rm GeV$. Excluding  the forward
and backward regions, the $\sigma (\gamma \gamma \to A^0A^0)$
cross-section is found in the $(0.1-0.2)\rm fb $ region.\par

At sufficiently high energies, both amplitudes $F_{++}$ and
$F_{+-}$ are found to be to largely imaginary; an effect
reminiscent, but not so predominant, as the one noticed in
neutral gauge boson production \cite{gg-gZ, ggZZ}.
On the contrary, nothing like this
appears for the $F_{++}$ amplitude in $\gamma \gamma \to
h^0 h^0$, in the Standard Model. It seems that the predominance of
the imaginary part of a loop amplitude at high energies,
is somehow associated with the predominance
of a $W$-involving loop.  The understanding of
such properties may  be useful for new
physics searches; since \eg~ for $\gamma \gamma \to \gamma
\gamma $ they determine the way the interference between
the "old" and possible forms of "new" physics may
appear \cite{extra-dim}.

Thus, after the discovery of $A^0$ and the study of the
single production process $\gamma \gamma \to A^0$
\cite{single-H0A0}, the study of the double $A^0$ production
through $\gamma \gamma \to A^0 A^0$,
should certainly be  useful for verifying the Higgs
identification.

\newpage

\renewcommand{\theequation}{A.\arabic{equation}}
\renewcommand{\thesection}{A.\arabic{section}}
\setcounter{equation}{0}
\setcounter{section}{0}

{\large \bf Appendix A: The  MSSM vertices for
$\gamma \gamma \to A^0 A^0 $.}

In order to reduce the number of diagrams contributing
to $\gamma \gamma \to A^0 A^0 $, we
use the  nonlinear gauge defined by  the gauge fixing term
\bqa
     \L_{GF} & = &
      -\frac{1}{\xi_W}F^+ F^- -\frac{1}{2\xi_Z}(F_Z)^2
      -\frac{1}{2\xi_\gamma}(F_\gamma)^2 ~,
  \nonumber \\
     F^\pm &=&~ \partial^\mu W^\pm_\mu
     \pm i\xi_W m_{W} G^\pm \pm i
     g^{\prime}B^{\mu}W^{\pm}_{\mu}  ~ , \nonumber \\
     F_Z  &=&~ \partial^\mu Z_\mu +\xi_Z m_{Z} G^0  ~ ,\nonumber \\
     F_\gamma&=&~ \partial^\mu A_\mu ~, \label{gauge-fixing-L}
\eqa
which is free from  $\gamma W^\pm G^\mp$ and $Z W^\pm G^\mp$
vertices \cite{Dicus}. The implied ghost-photon and ghost-scalar
field interactions then are
\begin{eqnarray}
\L_{FP}&=& ieA_{\mu}(\partial^{\mu}\bar{\eta}^- \eta^+
-\partial^{\mu}\bar{\eta}^+ \eta^- +\bar{\eta}^+
\partial^{\mu} \eta^- -\bar{\eta}^- \partial^{\mu}\eta^+)
+ e^2 A_{\mu}A^{\mu}(\bar{\eta}^+ \eta^- +\bar{\eta}^- \eta^{+})
 \nonumber \\
&-& {1\over 2}\xi_W gm_W
 (\bar{\eta}^+ \eta^- +\bar{\eta}^- \eta^{+})
[\cos(\alpha-\beta)H^0 +\sin(\beta- \alpha)h^0]
~ . \label{ghost-L}
\end{eqnarray}

\begin{figure}[p]
\vspace*{-2cm}
\[
\epsfig{file=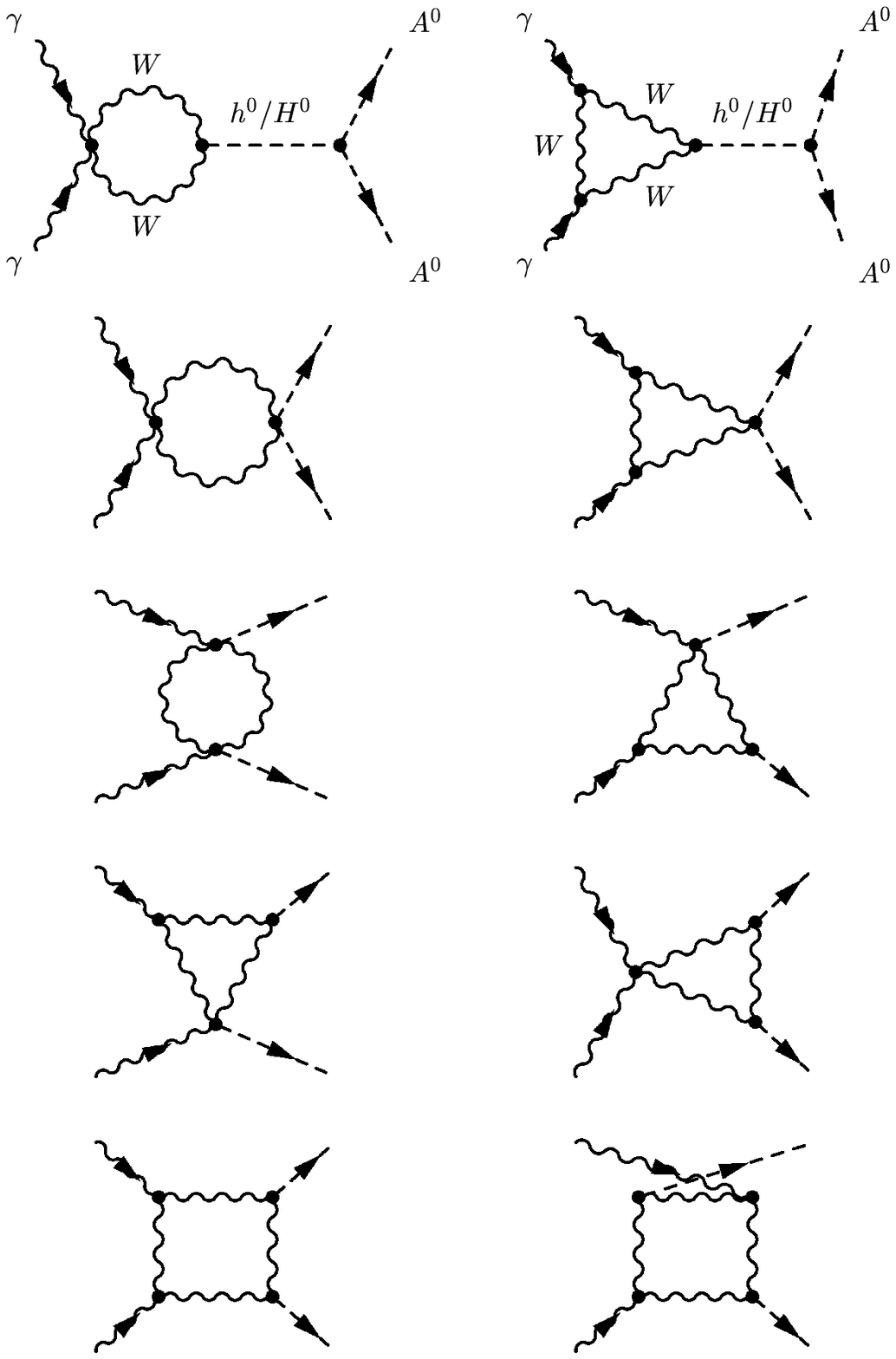,height=20cm,width=12cm}
\]
\caption[1]{Diagrams describing the $(W^\pm,~ H^\pm)$
loop contribution to $\gamma \gamma \to A^0 A^0$ in SUSY models.
 The internal wavy lines describe either a $W^\pm$
 propagator (together with the associated Goldstone and
ghost ones) or an   $H^\pm$ one. }
\label{WH-diag}
\end{figure}

\begin{figure}[p]
\vspace*{-3cm}
\[
\epsfig{file=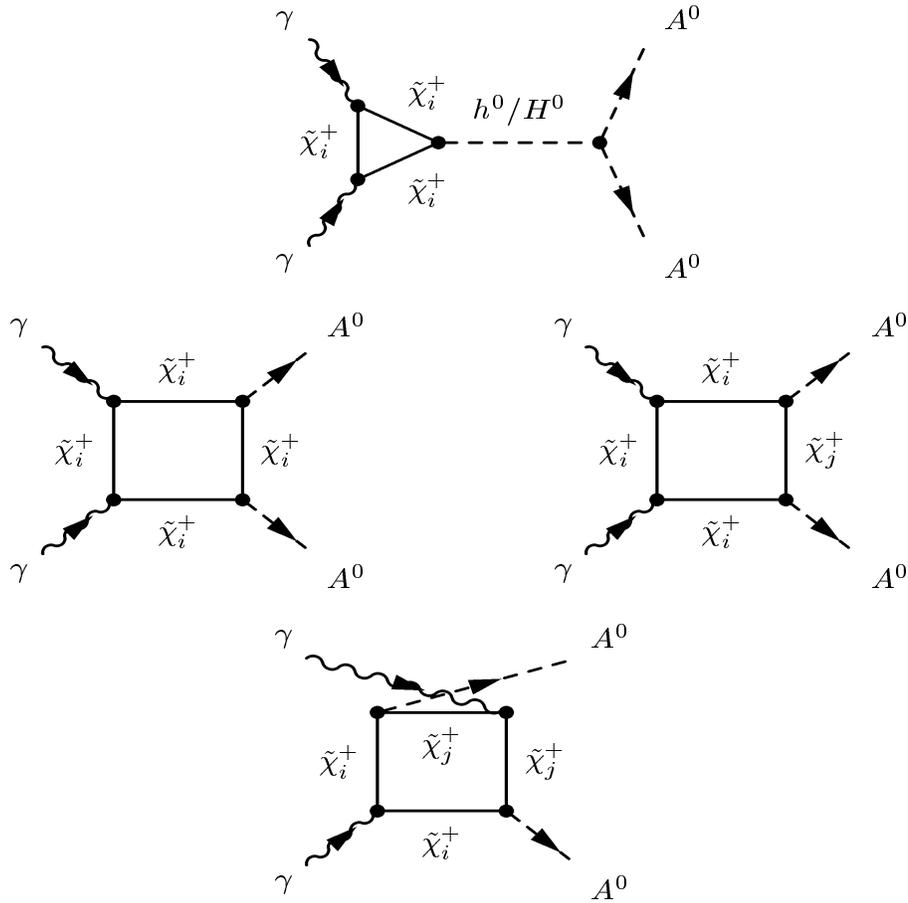,height=12cm,width=12cm}
\]
\caption[1]{Diagrams describing the chargino
contributions to the $\gamma \gamma \to A^0 A^0$ in
SUSY models. The last three boxes may involve both
$\tilde \chi_1 (\equiv \tilde \chi_1^+) $ and $\tilde \chi_2$,
running simultaneously  along the loop.}
\label{chargino-diag}
\end{figure}

\begin{figure}[hbt]
\vspace*{-2cm}
\[
\epsfig{file=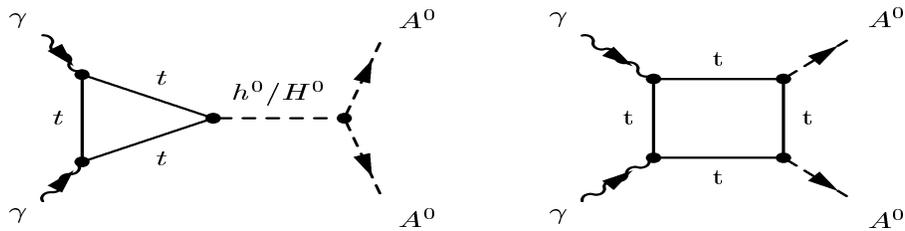,height=3cm,width=12cm}
\]
\caption[1]{Diagrams describing the  top
contributions to  $\gamma \gamma \to A^0 A^0$ in
SUSY models. Similarly for the $b$-quark loop. }
\label{top-diag}
\end{figure}

\begin{figure}[p]
\vspace*{-3cm}
\[
\epsfig{file=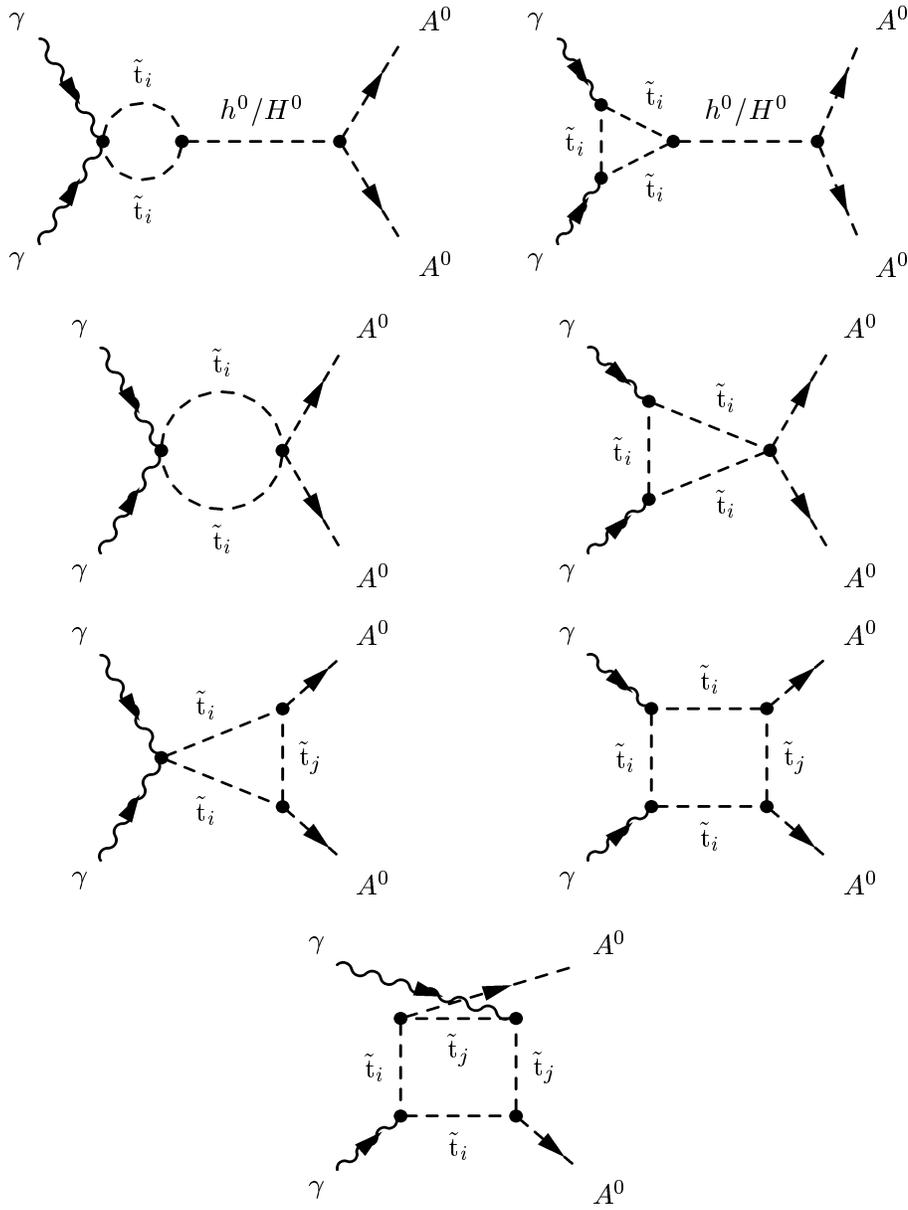,height=16cm,width=12cm}
\]
\caption[1]{The  stop contributions to
$\gamma \gamma \to A^0 A^0$ in SUSY models.
The last three diagrams may involve both $\tilde t_1$
and  $\tilde t_2$ running simultaneously  along the loop.}
\label{stop-diag}
\end{figure}

The complete list  of diagrams contributing to the process
$\gamma \gamma \to A^0A^0$  in the present  gauge,  appear in
Figs.\ref{WH-diag}-\ref{stop-diag}. Below we give the interaction
Lagrangian describing the vertices for these sets of diagrams.
\\ \par

The diagrams in Fig.\ref{WH-diag} describe the
\underline{$(H^\pm,~ W^\pm)$} loop
contribution (together of course with the accompanying ghost and
Goldstone ones).
The relevant vertices involve, in
addition to the gauge boson self-interactions
present in SM, also the triple and quartic vertices
\cite{SUSY-reviews}
\bqa
 \L_{VH} &= &-{g\over 2}[(A^0\partrl^\mu H^-)W_\mu^{+}
 +(A^0\partrl^\mu H^+)W_\mu^{-}]
 +i \frac{g m_W}{2} (A^0 H^-G^+ - A^0 H^+G^-)
\nonumber \\
&+ &gm_W
[\cos(\beta-\alpha)H^0 +\sin(\beta-\alpha)h^0] W_\mu^+W^{-\mu}
-i e (H^-\partrl^\mu H^+) A_\mu
\nonumber \\
& + &\frac{ g \mw}{2\cwd }  \cos 2\beta ~
[\sin(\alpha+\beta) h^0- \cos(\alpha+\beta) H^0 ] G^+G^{-}
\nonumber \\
& + &\frac{ g \mw}{4\cwd }  \cos 2\beta ~
[ \cos(\alpha+\beta) H^0 -\sin(\alpha+\beta) h^0] A^0A^0
 \nonumber \\
 &+&  g m_W \left [ \sin(\alpha-\beta)-\frac{\cos 2\beta}{2\cwd}
\sin(\alpha+\beta) \right ]h^0 H^+H^-
  \nonumber \\
 &-& g m_W \left [ \cos(\alpha-\beta)
-\frac{\cos 2\beta}{2\cwd} \cos(\alpha+\beta)\right ]H^0 H^+H^-
\nonumber \\
  & + &\frac{g^2}{4} \left [ W_\mu^+ W^{-\mu}
 -\left (1-\frac{\cos^2 2\beta}{2\cwd}\right ) G^+G^-
  -\frac{\cos^2 2\beta}{2\cwd }   H^+H^-\right ] A^{0}A^{0}
\nonumber \\
&+& i\frac{g e}{2} A^\mu A^0 [W^+_\mu H^- - W^-_\mu
H^+] +e^2 H^+H^- A_\mu A^\mu  ~. \label{gauge-Higgs-vertex}
\eqa
On the basis of this we define the $h^0$-couplings
\footnote{For the definition of the scalar sector mixing angles we
follow the standard notation of \eg~ \cite{Djouadi-SUSY-group}.}
\bqa
g_{h\bar \eta \eta}  \equiv
-~{1\over 2}\xi_W gm_W \sin(\beta- \alpha) & ~, ~ &
g_{hWW}  \equiv  g \mw \sin(\beta-\alpha) ~ , \nonumber \\
g_{hGG} \equiv \frac{g \mw}{2\cwd} \cos2\beta ~ \sin(\alpha+\beta)
& ~,~ &
g_{hAA} \equiv -~\frac{g \mw}{2\cwd} \cos2\beta ~ \sin(\alpha+\beta)
~ , \nonumber
\eqa
\bq
  g_{hH^+H^-}=g m_W \left
[\sin(\alpha-\beta)-\frac{\cos 2\beta}{2\cwd}
\sin(\alpha+\beta)\right ] ~ ,  \label{gauge-h0-couplings}
\eq
and  the $H^0$-couplings
\bqa
g_{H^0 \bar \eta \eta}  \equiv
-~{1\over 2}\xi_W gm_W \cos(\beta- \alpha) & ~, ~ &
g_{H^0WW}  \equiv  g \mw \cos(\beta-\alpha) ~ , \nonumber \\
g_{H^0GG} \equiv -~ \frac{g \mw}{2\cwd} \cos2\beta ~ \cos(\alpha+\beta)
& ~,~ &
g_{H^0AA} \equiv \frac{g \mw}{2\cwd} \cos2\beta ~ \cos(\alpha+\beta)
~ , \nonumber
\eqa
\bq
g_{H^0H^+H^-}= -g m_W \left
[\cos(\alpha-\beta)-\frac{\cos 2\beta}{2\cwd}
\cos(\alpha+\beta)\right ] ~ ,  \label{gauge-H0-couplings}
\eq
which  are used in Appendix B for expressing
the (Higgs-$W$) loop
contribution of the diagrams in Fig.\ref{WH-diag},
 as well as  the s-channel $(h^0,~ H^0)$-pole diagrams
contained  in Figs.\ref{chargino-diag}-\ref{stop-diag}. \\

The \underline{chargino} loop contribution is described
by the diagrams in
Fig.\ref{chargino-diag}. To define them  we first list
the chargino mass matrix  term   as\footnote{The
gaugino fields are defined so that
they satisfy $C \bar{\tilde W}^{+\tau} =\tilde W^-$. In such
a case there is no $i$ in front of $\tilde W^\pm$
in (\ref{chi-mass-matrix}).}
\bq
\L_{M_\chi} =-
\left ( \matrix{ \tilde W^{-\tau} , \tilde H_1^{-\tau}}
\right )_L  \cdot C \cdot \left (\matrix{
  M_2 & \sqrt{2} \mw \sin\beta \cr
 \sqrt{2} \mw \cos \beta   &  + \mu \cr } \right )
 \left (\matrix {\tilde W^+ \cr \tilde H_2^+}\right )_L
~+ {\rm h.c.} ~   . \label{chi-mass-matrix}
\eq
Assuming that in MSSM there no new sources of CP-violation, apart
from those already known in the Yukawa part of SM; we take the
quantities $(M_2,~ \mu)$ as real, but of
arbitrary sign.  $C$ is the usual charge conjugation
matrix, and the $\tau$ index indicates transposition of
the spinorial  field. In terms of
\bq
\tilde D \equiv  \left [
(M_2^2+\mu^2+ 2\mwd)^2- 4 (M_2\mu-\mwd \sin(2\beta))^2
\right ]^{1/2} ~ , \label{Dtilde}
\eq
the physical chargino masses are expressed as
\bq
m_{\tilde \chi_1, \tilde \chi_2}
=\frac{1}{\sqrt{2}} [M_2^2 +\mu^2 +2\mwd \mp \tilde D ]^{1/2}
~ . \label{chi-mass}
\eq
The mixing-angles $\phi_R, \phi_L$ in the
$(\tilde W^+, \tilde H_2^+)_L$ and
$(\tilde W^-, \tilde H_1^-)_L$ sectors respectively, are defined
  so that they always lie in the second quarter
\bq
\frac{\pi}{2} \leq \phi_L < \pi  ~~~~ , ~~~~
\frac{\pi}{2} \leq \phi_R  < \pi  ~~~ . \label{chi-angle-range}
\eq
They are written as
\bqa
\cos\phi_L &=& -~ \frac{1}{\sqrt{2\tilde D}}
[\tilde D-M_2^2+\mu^2 +2\mwd \cos 2\beta ]^{1/2} ~~ ,
\nonumber \\
\cos\phi_R &=& -~ \frac{1}{\sqrt{2\tilde D}}
[\tilde D-M_2^2+\mu^2 -2\mwd \cos 2\beta ]^{1/2} ~~ .
\label{chi-angles}
\eqa
We always describe the chargino field  so that it absorbs a
 positive chargino particle; \ie~
$\tchi_j \equiv {\tchi_j}^+$ $(j=1,2)$.
Using this and the sign-quantities
\bqa
\B_L & = &\Sn (\mu \sin\beta +M_2 \cos\beta)  ~,
\nonumber \\
\B_R & = & \Sn (\mu \cos\beta +M_2 \sin \beta) ~,
\nonumber \\
\Del_1 &= &
\Sn (M_2 [\tilde D-M_2^2+\mu^2-2\mwd] -2 \mwd \mu \sin 2\beta )  ~, ~
\nonumber \\
 \Del_2 &=&
\Sn (\mu [\tilde D-M_2^2+\mu^2 +2\mwd] +2 \mwd M_2 \sin 2\beta ) ~ ,
\nonumber \\
\B_{LR} & \equiv & \Sn \left(M_2 \mu+
\frac{\mu^2+M_2^2}{2}\sin 2\beta \right ) =\B_L \B_R ~ ,
\nonumber \\
\Del_{12} &\equiv & \Sn (M_2 \mu -\mwd \sin 2\beta)=
 \Del_1  \Del_2 ~ , \label{signs}
\eqa
the neutral gauge boson-chargino couplings are written as
\bqa
\L & = & -e A^\mu \bar{\tchi}_j \gamma_\mu \tchi_j
-{e\over 2s_Wc_W}  Z^\mu \bar{\tchi}_j \left  ( \gamma_\mu g_{vj}
- \gamma_\mu \gamma_5 g_{aj} \right ) \tchi_j
\nonumber \\
&& -~{e\over2s_Wc_W}  Z^\mu \left [\bar{\tchi}_1 \left  ( \gamma_\mu g_{v12}
- \gamma_\mu \gamma_5 g_{a12} \right ) \tchi_2 +
\mbox{h.c.} \right ]
  ~~ , \label{gauge-chi-vertex}
\eqa
with
\bqa
g_{v1}= {3\over2}-2s^2_W+{1\over4}[\cos2\phi_L+\cos2\phi_R] &~ , ~&
g_{a1}=- ~{1\over4}[\cos2\phi_L-\cos2\phi_R] ~ ,
\label{Zchi1-couplings} \\
g_{v2}= {3\over2}-2s^2_W-~{1\over4}[\cos2\phi_L+\cos2\phi_R] &~ , ~&
g_{a2}={1\over4}[\cos2\phi_L-\cos2\phi_R] ~ ,
\label{Zchi2-couplings}
\eqa
\bqa
&&g_{v12}=
-~{\Sn(M_2)\over4}
[\B_R ~ \Del_{12} \sin2\phi_R+ \B_L \sin2\phi_L] ~ ,
\nonumber\\
&&g_{a12}=-~{\Sn(M_2)\over4} [\B_R ~  \Del_{12}\sin2\phi_R -
\B_L \sin2\phi_L] ~ ,
\label{Zchi12-couplings}
\eqa
where the sign-factors $(\B_L ,~ \B_R, ~  \Del_{12})$
 are given
in\footnote{ These expressions are equivalent to those given \eg~
in \cite{neutral-model}, where a more common definition
of the $\phi_{L,R}$-angles is employed.} (\ref{signs}).
 \par

The corresponding  chargino-neutral Higgs vertices are
\bqa
\L_{A^{0}}& = & i A^{0}\Bigg [g_{A1} \bar{\tchi}_1 \gamma_{5}\tchi_1
+g_{A2} \bar{\tchi}_2 \gamma_{5} \tchi_2
 +  \bar{\tchi}_1 \left (g_{As12} +\gamma_5 g_{Ap12}  \right) \tchi_2
-\bar{\tchi}_2 \left ( g_{As12} - \gamma_5 g_{Ap12}  \right) \tchi_1
\Bigg ] \nonumber \\
& + &
( g_{h1} h^{0} +g_{H^01} H^0 ) \bar{\tchi}_1 \tchi_1 +
( g_{h2} h^{0} +g_{H^02} H^0)  \bar{\tchi}_2 \tchi_2  ~ ,
\label{Higgs-chi-vertex}
\eqa
where
\bqa
g_{h1} &=& -~\frac{g \Del_1}{\sqrt{2}}[-\B_L \salpha \cfR \sfL +
\B_R \calpha \sfR \cfL ] ~,  \nonumber \\
g_{H^01} &=& -~\frac{g \Del_1}{\sqrt{2}}[ \B_L \calpha \cfR \sfL +
\B_R \salpha \sfR \cfL ] ~,  \nonumber \\
g_{h2} &=& \frac{g \Del_2}{\sqrt{2}}[-\B_R \salpha \sfR \cfL +
\B_L \calpha \sfL \cfR ] ~,  \nonumber \\
g_{H^02} &=& \frac{g \Del_2}{\sqrt{2}}[ \B_R \calpha \sfR \cfL +
\B_L \salpha \cfR \sfL ] ~,  \nonumber \\
g_{A1} &=& -~ \frac{g \Del_1}{\sqrt{2}}[\B_L \sbeta \cfR \sfL +
\B_R \cbeta \sfR \cfL ] ~,  \nonumber \\
g_{A2} &=& \frac{g \Del_2}{\sqrt{2}}[\B_R \sbeta \sfR \cfL +
\B_L \cbeta \sfL \cfR ] ~, \nonumber \\
g_{As12} &=& \frac{g~  \Sn(M_2)}{2\sqrt{2}}
\Big [\B_{LR}( \Del_1 \cbeta -\Del_2 \sbeta) \sfR \sfL
\nonumber \\
& - &(\Del_1 \sbeta -\Del_2\cbeta) \cfL \cfR \Big ] ~ ,
\nonumber \\
g_{Ap12} &=& \frac{g ~ \Sn(M_2)}{2\sqrt{2}}
\Big [\B_{LR}( \Del_1 \cbeta +\Del_2 \sbeta) \sfR \sfL
\nonumber \\
& - &(\Del_1 \sbeta +\Del_2\cbeta) \cfL \cfR \Big ] ~ .
\label{Higgs-chi-couplings}
\eqa
The appearance in
(\ref{Zchi12-couplings}, \ref{Higgs-chi-couplings}) of the
sign-factors  defined in (\ref{signs}), guarantees
that the physical charginos always have positive masses;
irrespective of the signs
of the arbitrary real parameters  $M_2$ and $\mu$.
These signs are of course intimately related to the definition of
the chargino mixing angles employed in
(\ref{chi-angle-range}, \ref{chi-angles}).\\

We next turn to \underline{$t$ and $b$ quark}
loop contribution. The relevant diagrams for the
$t$-quark case are shown in Fig.\ref{top-diag}.
The necessary vertices are determined by
\bqa
\L_{t}& =& -eA_{\mu}[ Q_t  \bar{t}\gamma^{\mu}t
+ Q_b  \bar{b}\gamma^{\mu}b ]
 + i\frac{g }{2m_W}A^{0}[m_t \cot \beta ~\bar{t}\gamma_{5}t
+ m_b \tan \beta~ \bar{b}\gamma_{5}b ]
\nonumber \\
&& -\frac{g m_t }{2 m_W \sin \beta}[h^0\calpha  +H^0 \salpha ]
\bar{t}t
\nonumber \\
&& -\frac{g m_b }{2 m_W \cos \beta}[H^0\calpha  -h^0 \salpha ]
\bar{b}b ~, \label{top-vertex}
\eqa
where $Q_t, ~ Q_b$ are the
$t$ and $b$ quark charges.
The implied $t$ quark couplings are
\bqa
&& g_{h^0tt}  =  -\frac{g m_t }{2 m_W \sin \beta} ~\calpha
~~~~ , ~~~~
g_{H^0tt}  = -\frac{g m_t }{2 m_W \sin \beta} ~\salpha
~ , \nonumber \\
&& g_{Att} = \frac{g m_t}{2m_W}\cot \beta  ~,
\label{top-couplings}
\eqa
and correspondingly fro the $b$-couplings.   \\

Finally, for the \underline{stop} loop contribution,
the relevant interaction Lagrangian is
\bqa
\L_{\Stop}& = & -i e Q_{t} A_{\mu} \left
[(\Stop_{1}^{*}\partrl^{\mu}\Stop_{1})+
(\Stop_{2}^{*}\partrl^{\mu}\Stop_{2}) \right ]
+e^2 Q_{t}^2
A_{\mu}A^{\mu}(\Stop_{1}^{*}\Stop_{1}+\Stop_{2}^{*}\Stop_{2})
\nonumber \\
& + & i\frac{g m_{t}}{2 m_{W}}(A_{t}\cot \beta
+\mu) A^{0}[ \Stop^{*}_{L}\Stop_{R}- \Stop^{*}_{R}\Stop_{L}]
\nonumber \\
&-& \Big [ \frac{g m_{t}^2 \cos \alpha}{m_{W}
\sin \beta}-g_{Z}m_{Z}\sin(\alpha+\beta)
\left (T^{(3)}_t - Q_t\sw^2\right ) \Big ]
h^{0} \Stop_{L}^{*}\Stop_{L}
\nonumber \\
&-& \Bigl [ \frac{g m_{t}^2 \cos \alpha}{m_{W} \sin
\beta}-Q_t g_{Z}m_{Z}\sin(\alpha+\beta)\sw^2)
\Bigr ]h^{0} \Stop_{R}^{*}\Stop_{R}
\nonumber \\
&-& \frac{g m_{t}}{2 m_{W} \sin
\beta}(\mu \sin \alpha + A_{t}\cos \alpha)h^{0}
( \Stop_{R}^{*}\Stop_{L}+  \Stop_{L}^{*}\Stop_{R})
\nonumber \\
&-& \Bigl [ \frac{g m_{t}^2 \sin \alpha}{m_{W}
\sin \beta}+g_{Z}m_{Z}\cos(\alpha+\beta)
\left (T^{(3)}_t - Q_t \sw^2 \right )\Bigr]
H^{0} \Stop_{L}^{*}\Stop_{L}
\nonumber \\
&-& \Bigl [ \frac{g m_{t}^2 \sin \alpha}{m_{W} \sin
\beta}+Q_t g_{Z}m_{Z}\cos(\alpha+\beta)\sw^2)
\Bigr ] H^{0} \Stop_{R}^{*}\Stop_{R}
\nonumber \\
&+& \frac{g m_{t}}{2 m_{W} \sin
\beta}(\mu \cos \alpha - A_{t}\sin \alpha)H^{0}
( \Stop_{R}^{*}\Stop_{L}+ \Stop_{L}^{*}\Stop_{R})
\nonumber \\
&-& \Bigl [ \frac{g^2 m_{t}^2}{4m^{2}_{W}}\cot^{2}
\beta-\frac{1}{4}g^{2}_{Z}(T^{(3)}_t-\sw^2
Q_{t})\cos 2\beta \Bigr ] A^{0}A^{0} \Stop_{L}^{*}\Stop_{L}
\nonumber \\
&-& \Bigl [ \frac{g^2 m_{t}^2
}{4m^{2}_{W}}\cot^{2} \beta-\frac{1}{4}g^{2}_{Z}\sw^2 Q_{t} \cos
2\beta \Bigr ] A^{0}A^{0} \Stop_{R}^{*}\Stop_{R}
~ ~ ,\label{stop-vertex}
\end{eqnarray}
where, as usual, $g_Z=g/\cw$.
The various neutral Higgs-$\Stop_{L,R}$ couplings are determined
from the coefficients of the various terms
in (\ref{stop-vertex}). For two examples, we note\footnote{As usual,
  in the definition of the
$A^0A^0\Stop_L^*\Stop_L$ and  $A^0A^0\Stop_R^*\Stop_R$
couplings  from the last two terms in
(\ref{stop-vertex}), the relevant coefficient is multiplied by
a 2, due to the identity of the two $A^0$-fields.}
\[
g_{A\Stop_L\Stop_R}=\frac{g m_{t}}{2 m_{W}}(A_{t}\cot \beta
+\mu) ~ ,
\]
\[
g_{A A \Stop_R\Stop_R}= -2 \Bigl [ \frac{g^2 m_{t}^2
}{4m^{2}_{W}}\cot^{2} \beta-\frac{1}{4}g^{2}_{Z}\sw^2 Q_{t} \cos
2\beta \Bigr ] ~ .
\]
 For determining the corresponding couplings
to the physical $\Stop_{1,2}$ we write
\begin{equation}
    \llgm{\Stop_L \cr \Stop_R} \rrgm =
    \llgm{\cos\theta_t &  - \sin\theta_t \Sn(A_t-\mu \cot \beta) \cr
\sin\theta_t\Sn(A_t-\mu \cot \beta)   & \cos\theta_t}\rrgm
    \llgm{\Stop_1 \cr \Stop_2}\rrgm
~ , \label{stop-angle-matrix}
\end{equation}
where $\theta_t$ is fully determined by\footnote{ The quantities
$m^2_{\Stop_L}$, $m^2_{\Stop_R}$ in (\ref{stop-angles})
are the usual soft SUSY breaking
parameters in which the small D-contributions have also been
included.}
\bq
\sin(2\theta_t)=
\frac{2 m_t |A_t-\mu \cot\beta|}{m^2_{\Stop_1}-m^2_{\Stop_2}}
~~~~ , ~~ ~~~~~
\cos(2 \theta_t)=\frac{m^2_{\Stop_L}-m^2_{\Stop_R}}
{m^2_{\Stop_1}-m^2_{\Stop_2}} ~~~, \label{stop-angles}
\eq
while  $A_t$ is also real. We observe  from (\ref{stop-angles})
that
\[
\frac{\pi}{2} < \theta_t < \pi ~~~~~~ , ~
\]
since $m_{\Stop_1} <m_{\Stop_2}$, by definition.
We have checked that this
stop-mixing-formalism is equivalent to the usual one found \eg~
in \cite{SUSY-reviews, Djouadi-SUSY-group, stop-mixing, ggZZ}.
\par

\vspace{2cm}

\renewcommand{\theequation}{B.\arabic{equation}}
\renewcommand{\thesection}{B.\arabic{section}}
\setcounter{equation}{0}
\setcounter{section}{0}

{\large \bf Appendix B: The MSSM contributions to
 $\gamma \gamma \to A^0 A^0 $.}

The invariant helicity amplitudes for  the process
\begin{equation}
\gamma (p_1,\lambda_1) \gamma (p_2,\lambda_2) \to A^0(p_3)
A^0(p_4) \ \ , \label{process-ggAA}
\end{equation}
are denoted as\footnote{Their sign is related to the sign of
the $S$-matrix through   $S_{\lambda_1 \lambda_2}
= 1+i (2\pi)^4 \delta(p_f-p_i)
F_{\lambda_1 \lambda_2 }$. }
$F_{\lambda_1 \lambda_2 }(\s,\t,\u)$,
where the particle-momenta and
helicities of the incoming  photons,
 are indicated in parentheses.
Assuming no new (beyond SM) source of CP violation,
these invariant helicity amplitudes  satisfy
\bq
F_{\lambda_1, \lambda_2 }(\s,\t,\u)=
F_{-\lambda_1, -\lambda_2 }(\s,\t,\u) ~~~,
\label{CP-constraint}
\eq
which implies that there are  only two independent helicity
amplitudes; namely $F_{++}$ and $F_{+-}$.
As in \cite{ggZZ} we make  the definitions
\bqa
& \s=(p_1+p_2)^2=\frac{4 m_A^2}{1-\beta_A^2}
~~ ~,~ ~ \t=(p_1-p_3)^2 ~ ~,~ ~ \u=(p_1-p_4)^2 ~ & ,
\label{kin1} \\
& \s_4=\s-4 m_{A}^2 ~, ~ \s_2=\s-2 m_{A}^2 ~, ~ \t_1=\t- m_{A}^2 ~,~
\u_1=\u- m_{A}^2 ~ ~ & , \label{kin2} \\
& \t=\mad -\frac{\s}{2}(1-\beta_A \cos\vartheta^*) ~~ , ~~
\u=\mad -\frac{\s}{2}(1+\beta_A \cos\vartheta^*) &~ ,
\label{kin3} \\
 & Y=\t \u -m_A^4=\frac{\s^2\beta_A^2}{4}
\sin^2 \vartheta^* ~& . \label{kin4}
\eqa
where $\beta_A$ is the $A^0$-velocity in the $A^0A^0$-c.m. frame,
and $\vartheta^*$ the c.m. scattering angle.
Moreover, the combinations
\begin{equation}
m_{ab}^2=m_{A}^2+m_{a}^2-m_{b}^2 ~~,~~ \s_{ab}=\s-m_{ab}^2
~ , \label{kin5}
\end{equation}
often appear below for the charged particle  pairs
 $(a,b)=(H^\pm, W^\mp)$,
$(W^\mp, H^\pm)$, and  $(\tchi_1, \tchi_2)$.\par

All 1-loop results are expressed in terms
of  the $C_0$ and $D_0$ Passarino-Veltman functions
\cite{Passarino}, for which we follow the
notation of \cite{Hagiwara}. Similarly to \cite{ggZZ}, we
also introduce the short hand writing\footnote{In the middle terms
of (\ref{C0s}-\ref{DAAtu}) $k_1=p_1$, $k_2=p_2$ denote
the momenta of the photons, while  $k_3=-p_3$, $k_4=-p_4$
those of the $A^0$, always taken as incoming;
compare (\ref{process-ggAA}).}
\bqa
C_{0}^{abc}(\s)\equiv
 C_0(k_1, k_2)& = & C_{0}(0,0,\s; m_{a},m_{b},m_{c}) ~ ,
\label{C0s} \\
C_{A}^{abc}(\t)\equiv
C_0(k_3, k_1) &= & C_{0}(m_{A}^2,0,\t;m_{a},m_{b},m_{c})
~ ,\label{CAt} \\
C_{AA}^{abc}(\s) \equiv C_0(k_3, k_4) & = &
C_{0}(m_{A}^2,m_{A}^2,\s;m_{a},m_{b},m_{c})
~ , \label{CAAs} \\
D_{AA}^{abcd}(\s,\t) \equiv D_0(k_4, k_3, k_1)& = &
D_{0}(m_{A}^2,m_{A}^2,0,0,\s,\t;m_{a},m_{b},m_{c},m_{d})
~ , \label{DAAst}  \\
D_{AA}^{abcd}(\s,\u) \equiv D_0(k_3, k_4, k_1)& = &
D_{0}(m_{A}^2,m_{A}^2,0,0,\s,\u;m_{a},m_{b},m_{c},m_{d})
~ , \label{DAAsu}  \\
D_{AA}^{abcd}(\t,\u)\equiv D_0(k_3, k_1, k_4)
& = &D_{0}(m_{A}^2,0,m_{A}^2,0,\t,\u; m_{a},m_{b},m_{c},m_{d})
~ , \label{DAAtu} \\
D_{AA}^{abcd}(\u,\t)\equiv D_0(k_4, k_1, k_3)
& = &D_{0}(m_{A}^2,0,m_{A}^2,0,\u,\t; m_{a},m_{b},m_{c},m_{d})
~ , \label{DAAut}
\eqa
in which we have also emphasized the fact that the
masses running along the various sides of the loop may be
different.

The fact that the masses along the loops in
Figs.\ref{WH-diag}, \ref{chargino-diag}, \ref{stop-diag}
may be different, considerably complicates the formulae.
Nevertheless, expressions analogous to those encountered
for the SM contributions to $\gamma \gamma \to ZZ$ \cite{ggZZ}
may be defined, which allows writing  the amplitudes in a
compact way. We thus define
\bqa
\tilde F^{ab}(\s,\t,\u) &= &D_{AA}^{abba}(\t,\u)+D_{AA}^{abaa}(\s,\t)+
D_{AA}^{abaa}(\s,\u) ~ , \label{Fstu} \\
E_{1}^{ab}(\s,\t)& = & \t_1 \left[C_{A}^{abb}(\t)+  C_{A}^{baa}(\t)
\right ]-\s \t D_{AA}^{abaa}(\s,\t)
~ , \label{E1st} \\
E_{2}^{ab}(\t,\u)& = &\t_1 \left [ C_{A}^{abb}(\t)+C_{A}^{baa}(\t)
\right ]+ \u_1 \left [ C_{A}^{abb}(\u)+C_{A}^{baa}(\u) \right ]
\nonumber \\
&- &Y D_{AA}^{abba}(\t,\u) ~, \label{E2tu}
\eqa
which are closely related to  the definitions
in Eqs.(A.22-A.24) in \cite{ggZZ}.
We also  note that
\bqa
& D_{AA}^{abba}(\t,\u) = D_{AA}^{abba}(\u,\t) =
D_{AA}^{baab}(\t,\u)= D_{AA}^{baab}(\u,\t)
~~ , &~~  \nonumber  \\
& \tilde F^{ab}(\s,\t,\u)=\tilde F^{ab}(\s,\u,\t)
~~~, ~~~ E_{2}^{ab}(\t,\u) =E_{2}^{ab}(\u,\t) =
 E_{2}^{ba}(\t,\u)~~ . &
\label{D-E2-relations}
\eqa \\

\noindent
\underline{\bf The $(W^\pm, ~ H^\pm)$-loop diagrams.}\\
There two kinds of contributions to the  invariant amplitudes
$F_{\lambda_1\lambda_2}(\gamma \gamma \to A^0A^0)$ from the
diagrams of Fig.\ref{WH-diag}. The first arises
from the  two diagrams in the first row of Fig.\ref{WH-diag}
 and contains  $\s$-pole contributions generated by
exchanging the CP-even neutral Higgs particles $h^0$ and $H^0$.
For  the  $\s$-channel $h^0$-case, this  is given by
\bqa
&& F_{++}^{WH^\pm(h^0-\mbox{pole})}
= \frac{ e^2 g_{hAA}}{8 \pi^2 (\s-m_{h}^2)} \bigg \{
g_{hH^{+}H^{-}} \big [1 +2 m_{H^\pm}^2~ C_{0}^{H^+H^+H^+}(\s)
\big ]+ g_{hGG}-2g_{h\eta\eta}\nonumber \\
&&
-4g_{hWW} +  2 \left [(g_{hGG}-2g_{h\eta\eta}-4g_{hWW})
 m_{W}^2+2 g_{hWW} \s \right ] C_{0}^{WWW}(\s) \bigg \} ~ ,
\label{FppWH-h-pole}
\eqa
while for the  $H^0$-case we get
\begin{equation}
F_{++}^{WH^\pm(H^0-\mbox{pole})}=F_{++}^{WH^\pm(h^0-\mbox{pole})}
( h^0 \rightarrow H^0) ~~ , \label{FppWH-H-pole}
\end{equation}
where  the  needed $h^0$ and $H^0$ couplings are given in
(\ref{gauge-h0-couplings}, \ref{gauge-H0-couplings}).
For such contributions we obviously also have
\begin{equation}
F_{+-}^{WH^\pm(H^0-\mbox{pole})}=
F_{+-}^{WH^\pm(h^0-\mbox{pole})}= 0
 ~~ . \label{FpmWH-pole}
\end{equation}

The second comes   from the 3rd to last
diagrams in Fig.\ref{WH-diag}, and it is written as
\begin{eqnarray}
&& F_{++}^{WH^\pm}=\frac{e^2 g^2}{16\pi^2 } \Bigg \{ 4 +
\left [6-\frac{\cos^2(2\beta)}{\cwd}
\right ] m_{W}^2 C_{0}^{WWW}(\s)+ \left
[2+\frac{\cos^2 (2\beta)}{\cwd} \right ] m_{H^\pm}^2
C_{0}^{H^\pm H^\pm H^\pm }(\s)
\nonumber \\
&&+ 2 \frac{m_{H W}^2}{\s}
E_{2}^{H^\pm W}(\t,\u)+ 2 m_{H^\pm}^2 \s_{HW}
\tilde F^{H^\pm W}(\s,\t,\u)\nonumber \\
&& +  2 (\s m_{H^\pm}^2-m_{HW}^2
 m_{W}^2) \tilde F^{WH^\pm}(\s,\t,\u) \Bigg \} ~~,
\label{FppWH} \\
&& F_{+-}^{WH^\pm}= \frac{e^2 g^2}{16\pi^2 Y}
\Bigg \{ \s \left[2(m_{HW}^2
m_{WH}^2-Y)+\s (m_{H}^2-m_{W}^2)+\t \t_1+\u \u_1 \right ]
C_{0}^{WWW}(\s)
\nonumber \\
&& +
\s \s_{HW} (\s_{HW} -  m_{HW}^2)
C_{0}^{H^\pm H^\pm H^\pm}(\s)+\s_{H W}
(\t^2+\u^2-2 m_{A}^4) \left [
C_{AA}^{H^\pm WH^\pm}(\s)+C_{AA}^{WH^\pm W}(\s) \right ]
\nonumber \\
&&+ 2 \left [\s \s_{HW}(m_{H^\pm}^2-m_{W}^2)^2+Y \left [2\s
m_{W}^2+Y-m_{HW}^2(m_{W}^2+m_{H^\pm}^2) \right ]\right ]
\tilde F^{WH^\pm}(\s,\t,\u)
\nonumber \\
&& + \bigg \{ \s_{H W}
\left [\s (m_{H^\pm}^2-m_{W}^2)^2+\s\t (\t-2 m_{W}^2)
-2 m_{H^\pm}^2
\t_{1}^2 \right ]
\left [ D_{AA}^{H^\pm WH^\pm H^\pm}(\s,\t)-
D_{AA}^{WH^\pm WW}(\s,\t) \right ]
\nonumber \\
&&+ 2(m_{A}^4+\t^2-\t m_{WH}^2)E_{1}^{WH^\pm}(\s,\t)
+ (\t \leftrightarrow \u) \bigg \} \Bigg \} ~ , \label{FpmWH}
\end{eqnarray}
where all needed  quantities have been defined in
(\ref{kin1}-\ref{E2tu}).\\

\noindent
\underline{\bf The chargino loop diagrams}\\
The relevant diagrams are presented in Fig.\ref{chargino-diag}.
The first  diagram in Fig.\ref{chargino-diag}
 contains an $\s$-channel pole
 due to  $(h^0,~ H^0)$ exchanges, and
is characterized by  a single chargino
$\cxi$ running along the loop. It  gives
\begin{eqnarray}
F_{++}^{\cxi(\mbox{pole})}
=-~\frac{ e^2 m_{\cxi}}{4\pi^2} \biggl (\frac{g_{hi}
g_{hAA}}{\s-m_{h}^2}+\frac{g_{Hi} g_{H^0 AA}}{\s-m_{H}^2}\biggr )
\left [2+(4m_{\cxi}^2-\s)C_{0}^{\cxi\cxi\cxi}(\s)\right ]
~ , \label{Fppchi-pole}
\end{eqnarray}
\begin{eqnarray}
F_{+-}^{\cxi(\mbox{pole})}=0 ~ . \label{Fpmchi-pole}
\end{eqnarray}
The other  diagrams in Fig.\ref{chargino-diag}
involve contributions containing either a single chargino
running along the loop, or mixed contributions
where both charginos run. The single chargino contribution
 is
\begin{eqnarray}
F_{++}^{\cxi}& = &-~\frac{e^2 g_{Ai}^2}{4\pi^2}
\bigg \{2+4 m_{\cxi}^2
C_{0}^{\cxi\cxi\cxi}(\s) - m_{\cxi}^2
(\t+\u)\tilde F^{\cxi\cxi}(\s,\t,\u)
\nonumber \\
 &+& \frac{m_{A}^2}{\s} E_{2}^{\cxi\cxi}(\t,\u) \bigg \}
~ , \label{Fppchi} \\
F_{+-}^{\cxi}&=&-~ \frac{e^2 g_{Ai}^2}{8 \pi^2 Y}
\bigg \{\s(\s_{2}^2-2 Y)
C_{0}^{\cxi\cxi\cxi}(\s) + \s_2 (\t^2+\u^2-2
m_{A}^4)C_{AA}^{\cxi\cxi\cxi}(\s)
\nonumber \\
 &+& 2 m_{\cxi}^2 \s_2 Y \tilde F^{\cxi\cxi}(\s,\t,\u)+
 (\t^2+m_{A}^4)E_{1}^{\cxi\cxi}(\s,\t)
\nonumber \\
 &+&(\u^2+m_{A}^{4})E_{1}^{\cxi\cxi}(\s,\u) \bigg \}
~ . \label{Fpmchi}
\end{eqnarray}
Of course, in calculating the total "single"
chargino contribution, the
results in (\ref{Fppchi-pole}-\ref{Fpmchi}) should be summed for
both the $\tchi_1$ and $\tchi_2$ charginos. The necessary
couplings are given in (\ref{Higgs-chi-couplings}).

The considerably more complicated
mixed chargino  contribution, arising from the 
3rd and 4th diagram in Fig.\ref{chargino-diag}, is
\begin{eqnarray}
F_{++}^{\xo\xt}&=&- ~\frac{ e^2}{4 \pi^2}
(g_{As12}^2+g_{Ap12}^2) \bigg \{2+4
m_{\xo}^{2}C_{0}^{\xo\xo\xo}(\s)+\frac{1}{2\s}
(\s-X)E_{2}^{\xo\xt}(\t,\u)
\nonumber \\
&& - m_{\xo}\left
[\s \left (m_{\xo}+
\frac{(g_{As12}^2-g_{Ap12}^2)}{(g_{As12}^2+g_{Ap12}^2)}
~m_{\xt} \right )-m_{\xo}X \right ]
\tilde F^{\xo\xt}(\s,\t,\u)
\nonumber \\
&+ &(\xo \leftrightarrow \xt) \bigg \} ~,
\label{Fppchi1chi2} \\
F_{+-}^{\xo\xt}&=&- ~\frac{ e^2 (g_{As12}^2+g_{Ap12}^2)}{8\pi^2 Y}
\Bigg \{
\s\left [X(\s_{\xo\xt}-m_{\xo\xt})-2Y\right ]
C_{0}^{\xo\xo\xo}(\s)
\nonumber \\
&+& X \left
[(m_{\xo}^2+m_{\xt}^2)Y+\s (m_{\xo}^2-m_{\xt}^2)^2
\right] \tilde F^{\xo\xt}(\s,\t,\u)
\nonumber \\ &+&
X(\t^2+\u^2-2m_{A}^4)C_{AA}^{\xo\xt\xo}(\s)-(\t
X+Y)E_{1}^{\xo\xt}(\s,\t)\nonumber \\ &-&(\u
X+Y)E_{1}^{\xo\xt}(\s,\u) -(m_{\xo}^2-m_{\xt}^2)\left[2 \t_{1}^2
X+ Y(X-\s)\right] D_{AA}^{\xo\xt\xo\xo}(\s,\t)\nonumber\\ &-&
(m_{\xo}^2-m_{\xt}^2)\left[2 \u_{1}^2 X+ Y(X-\s)\right]
D_{AA}^{\xo\xt\xo\xo}(\s,\u) + (\xo \leftrightarrow \xt)\Bigg \}
~ , \label{Fpmchi1chi2}
\end{eqnarray}
which
where
\begin{equation}
X=\s_{2}+2m_{\xo}^2+2m_{\xt}^2+4m_{\xo}m_{\xt}~
\frac{(g_{As12}^2-g_{Ap12}^2)}{(g_{As12}^2+g_{Ap12}^2)} ~~,~~
\end{equation}
and the necessary couplings are given in
(\ref{Higgs-chi-couplings}).\\

\noindent
\underline{\bf The $t$ and $b$-quark loop diagrams.}\\
The top-loop contribution arises from the diagrams in
Fig.\ref{top-diag}.
The first of them contains the $(h^0,~H^0)$-pole contribution
\begin{eqnarray}
F_{++}^{t(\mbox{pole})}=-
\frac{3 e^2 Q_t^2 m_{t}}{4\pi^2 } \biggl (\frac{g_{htt} g_{hAA}}{
\s-m_{h}^2}+\frac{g_{H^0tt} g_{H^0AA}}{ \s-m_{H}^2} \biggr )
\left
[2+(4m_{t}^2-\s)C_{0}^{ttt}(\s)\right ] ~, \label{Fpptop-pole}
\end{eqnarray}
\begin{eqnarray}
F_{+-}^{t(\mbox{pole})}=0 ~~ , \label{Fpmtop-pole}
\end{eqnarray}
while the second gives
\begin{eqnarray}
F_{++}^{t}&= &- \frac{3 e^2 Q_t^2 g_{Att}^2}{4\pi^2}
\Big [2+4m_{t}^2 C_{0}^{ttt}(\s)-m_{t}^2
(\t+\u)\tilde F^{tt}(\s,\t,\u) \nonumber \\
&+ & \frac{m_{A}^2}{\s}E_{2}^{tt}(\t,\u)
\Big ]~  , \label{Fpptop} \\
F_{+-}^{t}&=&-~ \frac{3 e^2 Q_t^2 g_{Att}^2}{8\pi^2 Y}
\bigg \{ \s(\s_{2}^2-2Y)
C_{0}^{ttt}(\s)+\s_2(\t^2+\u^2-2m_{A}^4)C_{AA}^{ttt}(\s)
\nonumber \\
&+&
2m_{t}^2 \s_2 Y
\tilde F^{tt}(\s,\t,\u)+(\t^2+m_{A}^4)E_{1}^{tt}(\s,\t)+
(\u^2+m_{A}^4)E_{1}^{tt}(\s,\u) \bigg \} ~. \label{Fpmtop}
\end{eqnarray}
All needed couplings are given in (\ref{gauge-h0-couplings},
\ref{gauge-H0-couplings}, \ref{top-couplings}).
In (\ref{Fpptop-pole}-(\ref{Fpmtop}) a factor three
for colour has already been introduced. The corresponding
$b$-quark contribution is analogously obtained through
(\ref{top-vertex}) and the use of $Q_b$ instead of $Q_t$.\\

\noindent
\underline{\bf  $\Stop$-loop diagrams}\\
These diagrams are shown in Fig.\ref{stop-diag} and will be
relevant in case one or two stop sqarks turn out to be not too
heavy. The first two of these diagrams describe the
$(h^0,~ H^0)$ $\s$-channel pole
contributions and have just one kind of $\Stop_i$
running along the loop.  For each such $\Stop_i$,
the pole contribution is
\begin{eqnarray}
F_{++}^{\ti(\mbox{pole})}= \frac{3 e^2 Q_{t}^2}{8\pi^2}
\biggl( \frac{
g_{hAA}g_{h\ti\ti}}{\s-m_{h}^2}+\frac{
g_{HAA}g_{H\ti\ti}}{\s-m_{H}^2} \biggr ) \left [1+2m_{\ti}^2
C_{0}^{\ti\ti\ti}(\s)\right ] ~, \label{Fppstop-pole}
\end{eqnarray}
\begin{equation}
F_{+-}^{\ti(\mbox{pole})}=0 ~ ~ . \label{Fpmstop-loop}
\end{equation}
In addition, we have the loop contribution from the no-pole
last five diagrams of Fig.\ref{stop-diag}
\begin{eqnarray}
 F_{++}^{\ton\tt}& = & -~\frac{3 e^2 Q_{t}^2}{8\pi^2}
\Bigg \{g_{AA\ton\ton}
\left [1+2
m_{\ton}^2 C_{0}^{\ton\ton\ton}(\s)\right
] +\frac{g_{A\ton\tt}^2}{\s} E_{2}^{\ton\tt}(\t,\u)
\nonumber \\
&- & 2g_{A\ton\tt}^2 m_{\ton}^2 \tilde F^{\ton\tt}(\s,\t,\u)+
(\ton \leftrightarrow \tt) \Bigg \} ~, \label{Fppstop} \\
 F_{+-}^{\ton\tt} &= &
\frac{3 e^2 Q_{t}^2 g_{A\ton\tt}^2}{8\pi^2Y} \Bigg
\{\s(\s-2m_{\ton\tt}^2)C_{0}^{\ton\ton\ton}(\s)+
(\t^2+\u^2-2m_{A}^4) C_{AA}^{\ton\tt\ton}(\s)
\nonumber \\
& + &
Y(m_{\ton}^2+m_{\tt}^2)D_{AA}^{\ton\tt\tt\ton}(\t,\u)+
\s(m_{\ton}^2-m_{\tt}^2)^2 \tilde F^{\ton\tt}(\s,\t,\u)
\nonumber \\
& - & \t E_{1}^{\ton\tt}(\s,\t)-\u E_{1}^{\ton\tt}(\s,\u)
- 2(\s\t m_{\tt}^2+\t_{1}^2
m_{\ton}^2)D_{AA}^{\ton\tt\ton\ton}(\s,\t)
\nonumber \\
& - &2(\s\u
m_{\tt}^2+\u_{1}^2
m_{\ton}^2)D_{AA}^{\ton\tt\ton\ton}(\s,\u)
~~~~ + (\ton \leftrightarrow \tt) \Bigg \} ~~~~~ ,
 \label{Fpmstop}
\end{eqnarray}
which involve contributions either from a single $\Stop_j$
running along the loop,
or  mixed contributions involving both
both $\Stop_1,~ \Stop_2$.
In (\ref{Fppstop-pole}-(\ref{Fpmstop}) a factor three
for colour has already been introduced, while
the necessary couplings are given  by
combining (\ref{stop-vertex}, \ref{stop-angle-matrix}).\par

If other kinds of  sfermions turn out also to be light,
then their contribution
can readily be derived from (\ref{Fppstop-pole}-\ref{Fpmstop})
by changing  the appropriate couplings.

\clearpage
\newpage

\clearpage
\newpage

\clearpage
\newpage

\clearpage
\newpage

\end{document}